%% file: SSched_rsj2012.tex
\documentclass[onecolumn]{IEEEtran}

\input{crenv}

\graphicspath{{figuresTAC/}}

\begin{document}

\title{Design of State-based Schedulers for a \\
    Network of Control Loops}

\author{Chithrupa~Ramesh,~\IEEEmembership{Student~Member,~IEEE}
        Henrik~Sandberg,~\IEEEmembership{Member,~IEEE,}
        and~Karl~H.~Johansson,~\IEEEmembership{Senior~Member,~IEEE}% <-this % stops a space
%\thanks{This work was supported by the Swedish Research Council,
%        VINNOVA (The Swedish Governmental Agency for Innovation
%        Systems), the Swedish Foundation for Strategic Research,
%        the Knut and Alice Wallenberg Foundation and the EU
%        project FeedNetBack.}% <-this % stops a space
\thanks{C. Ramesh, H. Sandberg and Karl H. Johansson are with the ACCESS Linnaeus Centre,
Electrical Engineering, KTH Royal Institute of Technology, Stockholm,
Sweden. e-mail: (\{cramesh,hsan,kallej\}@ee.kth.se).}% <-this % stops a space
%\thanks{Manuscript received April 19, 2011; revised .}
}

%\markboth{Journal of \LaTeX\ Class Files,~Vol.~6, No.~1, January~2007}%
%{Ramesh \MakeLowercase{\textit{et al.}}: Design of State-based Schedulers for a Network of Control Loops}


\maketitle

\begin{abstract}
For a closed-loop system, which has a contention-based multiple access network on its sensor link, the Medium Access Controller (MAC) may discard some packets when the traffic on the link is high. We use a local state-based scheduler to select a few critical data packets to send to the MAC. In this paper, we analyze the impact of such a scheduler on the closed-loop system in the presence of traffic, and show that there is a dual effect with state-based scheduling. In general, this makes the optimal scheduler and controller hard to find. However, by removing past controls from the scheduling criterion, we find that certainty equivalence holds. This condition is related to the classical result of Bar-Shalom and Tse, and it leads to the design of a scheduler with a certainty equivalent controller. This design, however, does not result in an equivalent system to the original problem, in the sense of Witsenhausen. Computing the estimate is difficult, but can be simplified by introducing a symmetry constraint on the scheduler. Based on these findings, we propose a dual predictor architecture for the closed-loop system, which ensures separation between scheduler, observer and controller. We present an example of this architecture, which illustrates a network-aware event-triggering mechanism.
\end{abstract}


% \fi
\IEEEpeerreviewmaketitle

\section{Introduction}

\begin{figure}[!t]
\begin{center}
\includegraphics[scale=0.75]{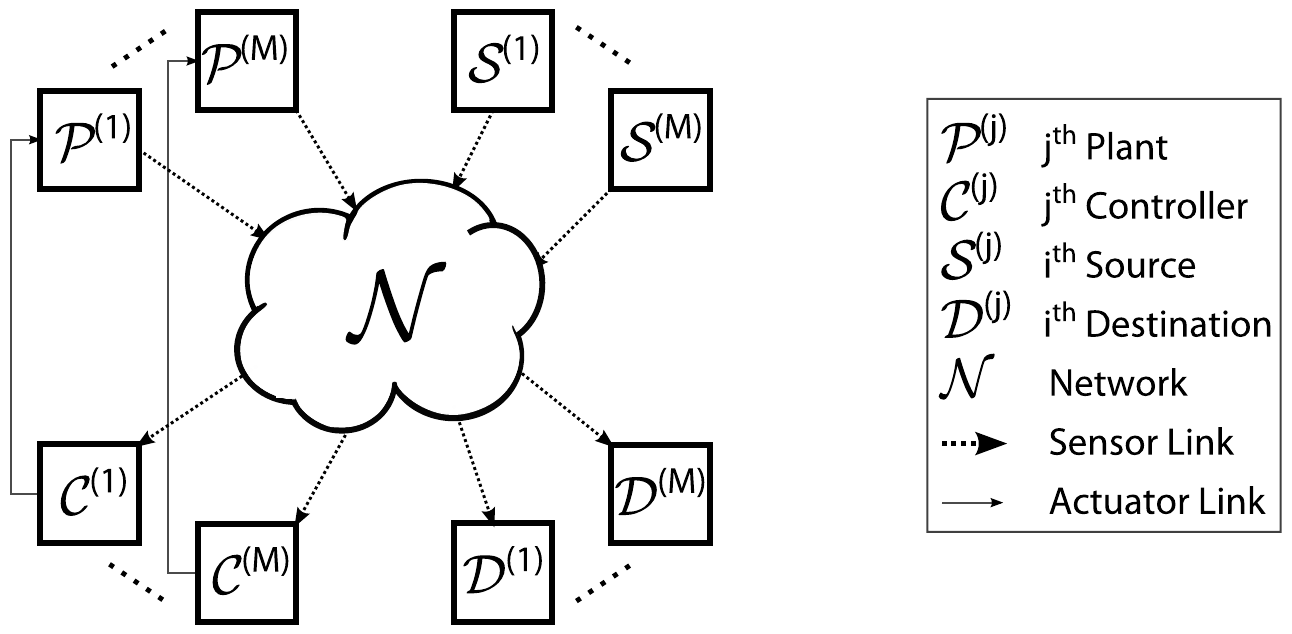}
\caption[NCSs with a shared sensing link]{A network of $M$ control loops, with each loop consisting of a plant $\mathcal{P}^{_{(j)}}$ and a controller $\mathcal{C}^{_{(j)}}$ for $j \in \{1,\dots,M\}$. The loops share access to a common medium on the sensor link, along with $N$ other communication flows from generic source-destination pairs. The controllers and actuators communicate over dedicated networks, not shared links.}
\label{Fig:macNCS}
\vspace{-5mm}
\end{center}
\end{figure}

\IEEEPARstart{C}{onsider} a network of control systems, where the communication between the individual sensors and controllers of different control loops occurs through a shared network, as shown in Fig. \ref{Fig:macNCS}. This is an important scenario, in the context of wireless Networked Control Systems (NCS), for industrial and process control \cite{Willig2005}. A medium access control layer is required in the sensor's protocol stack to arbitrate access to the shared network. To focus on the implications of a Medium Access Controller (MAC) on the sensor link, we assume that the communication between the controllers and the corresponding actuators occurs over a point-to-point network, not a shared network. This is a common architecture, in practice \cite{Xu2004,Otanez2002}. The MAC can implement a contention-free or a contention-based multiple access method, both of which have their own challenges \cite{Rom1990}. A contention-free multiple access method requires a dynamic scheduler to prevent poor channel utilization, and such a scheduler is hard to construct and implement on an interference-constrained shared network \cite{Ramaswami1989,Goldsmith2002}. Contention-based methods have proven popular in standards such as IEEE 802.15.4 \cite{Pollin2008}, as they facilitate an easy deployment on sensor nodes. However, such methods result in \emph{random access}, which could significantly deteriorate the performance of a closed-loop system \cite{Liu2004}. Thus, the design of a MAC for networked control systems is a challenging problem, and calls for innovative solutions \cite{Willig2008}.

In this paper, we explore the design of a \emph{state-aware} contention-based MAC, as opposed to an \emph{agnostic} contention-based MAC. The state-aware MAC is capable of influencing the randomness of channel access in favour of the state of the plant in the closed-loop system. However, directly using the state of the plant to determine an access probability may result in a MAC that is difficult to implement and analyze \cite{Bianchi2005}. Instead, we use the state of the plant to select packets to send to the MAC, motivated by an understanding of the two roles played by a MAC: Any random access method works by resolving contention between simultaneous channel access requests, thus spreading traffic that arrives in bursts. The carrier sense multiple access with collision avoidance (CSMA/CA) method does this by assigning a random back-off to packets that attempt to access a busy channel, thus spreading the traffic over a longer interval of time. Similarly, the p-persistent CSMA method does this by probabilistically limiting access to the channel and permitting a number of retransmissions if the channel is busy \cite{Kleinrock1975}. However, all of these methods permit only a finite number of retransmissions, beyond which the packet is discarded. We appropriate this latter role of discarding packets to a local \emph{state-based scheduler}, which sends fewer, but more important packets to the MAC for transmission across the network.

A similar strategy has previously been proposed from the more general perspective of reducing network traffic \cite{Otanez2002}. When applied to the newly posed NCS problem \cite{Hespanha2007}, this approach has driven the design of event-based sampling systems \cite{Rabi2006,Tabuada2007}, which have been shown to outperform periodically sampled systems under certain conditions \cite{AAstrom2002,Cervin2008,Rabi2009}. We approach the same problem from a different perspective, but one that leads to a \emph{network-aware} design of event triggering methods.

There are two main contributions in this paper. The first contribution is an analysis of the impact of having a state-based scheduler in the closed-loop. Primarily, a state-based scheduler permits the information available at the controller to be altered with the plant state. This information is not entirely random, like in the case of packet losses due to a noisy channel \cite{Schenato2007,Gupta2007}, and it can result in a sharply asymmetrical estimation error, unlike in the case of encoder design over limited data rate channels \cite{Tatikonda2004,Nair2007}. It seems reasonable to ask if we can use the controller to move the plant state across the threshold and force a transmission? If this is possible, the controller plays two roles: the first one being to control the plant, and the second one being to control the information available at the next time step. This relates to the classical concept of a dual effect, as described in \cite{Feldbaum1961,AAstrom1995}. The answer to this question determines the ease of optimal controller design, as the Certainty Equivalence Principle would not hold if there is a dual effect \cite{Bar-Shalom1974}. We examine our system and find that there is a dual effect with a state-based scheduler in the closed-loop, and thus, that the certainty equivalence principle does not hold. Hence, the optimal state-based scheduler, estimator and controller designs are coupled. A restriction on the input arguments to the state-based scheduler, such that these arguments are no longer a function of the past control actions, renders the setup free of a dual effect, and enables the certainty equivalence principle to hold. These results can be seen as an interpretation, within the state-based scheduler setup, of the classical work on information patterns \cite{Ho1980}, dual effect, certainty equivalence and separation by Witsenhausen \cite{Witsenhausen1971}, Bar-Shalom and Tse \cite{Bar-Shalom1974}, and on adaptive control by Feldbaum \cite{Feldbaum1961}, {\AA}str\"{o}m and Wittenmark \cite{AAstrom1995} and others \cite{Filatov2000}.

The second contribution of this paper is on the \emph{dual predictor architecture}, which is our proposed solution to the state-based scheduler design problem. In this architecture, the state-based scheduler thresholds the squared difference of the innovation contained in the latest measurement to the estimator across the network. This results in an optimal certainty equivalent controller, and a simple observer which generates the minimum mean-squared error (MMSE) estimate. Tuning parameters in the state-based scheduler in this architecture based on the current network traffic could result in a scheduling law that guarantees a probabilistic performance. This is not easy to show, in general, as the performance analysis of a closed-loop system with a state-based scheduler in a multiple access network is a difficult problem \cite{Cervin2008,Rabi2009}. However, we illustrate the guaranteed performance using simulations, and thus claim that the state-based scheduler we propose results in a network-aware event-triggering mechanism.

The rest of the paper is organized as follows. In Section~\ref{S:Prelim}, we present the problem formulation. In Section~\ref{S:SSched}, we derive theoretical results for the case when full state information is available, with and without exogenous network traffic. In Section~\ref{S:CSArch}, we present the dual predictor architecture. We look at an extension to output-based systems in Section~\ref{S:ExtnDisc}. We present an example, which illustrates our notion of network-aware event-triggering, in Section~\ref{S:Examples}. Providing performance guarantees remains a difficult problem, as we indicate under future work, along with the conclusions, in Section~\ref{S:ConclFW}.

\section{Preliminaries} \label{S:Prelim}
\IEEEPARstart{W}{e} present the problem setup and a few important definitions, along with a review of the classical concepts of dual effect and certainty equivalence in this section.

\subsection{Problem Formulation} \label{S:ProbSetup4}

\begin{figure}[t!]
\begin{center}
\hspace{-5mm}
\subfigure[A multiple access (MA) scenario for NCSs]{\label{Fig:SystemModel} \includegraphics[height=5cm,width=8cm,keepaspectratio=true]{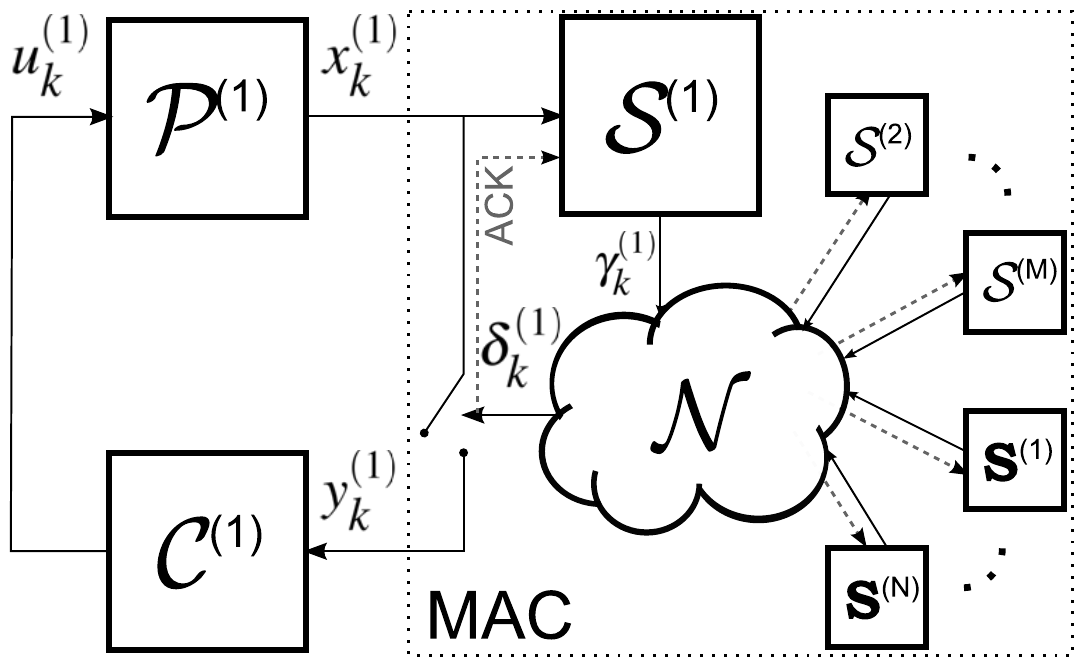}} \hspace{1cm}
\subfigure[The MA model for each closed-loop system]{\label{Fig:MAmodel} \includegraphics[height=5cm,width=7cm,keepaspectratio=true]{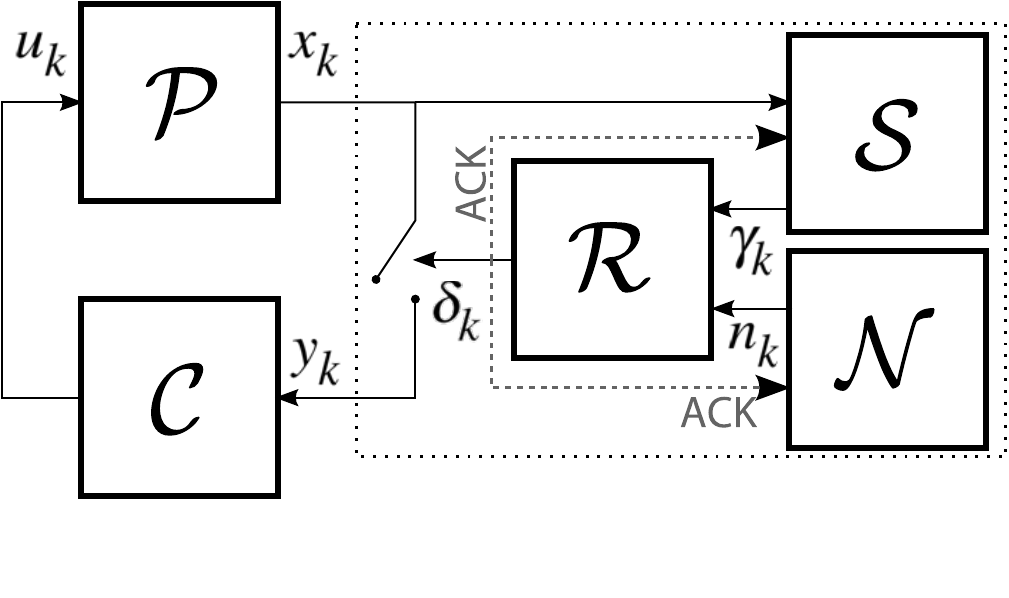}}
\caption[State-aware MAC model for heterogenous networks]{{\small A plant ($\mathcal{P}^{{(1)}}$), state-based scheduler ($\mathcal{S}^{{(1)}}$) and controller ($\mathcal{C}^{{(1)}}$) share the network ($\mathcal{N}$) with $M-1$ other closed-loop systems with state-based schedulers ($\mathcal{S}^{{(j)}}, j \in \{2,\dots,M\} $), and $N$ generic sources ($\mathbf{S}^{{(i)}}, i \in \{1,\dots,N\} $), in (a). A model, from the perspective of a closed-loop system in the network, is depicted in (b).}}
\end{center}
\vspace{-5mm}
\end{figure}

We consider a network of $M$ control loops, as shown in Fig.~\ref{Fig:SystemModel}. Each control loop, for $j \in \{1,\dots,M\}$, consists of a plant $\mathcal{P}^{_{(j)}}$, a state-based scheduler $\mathcal{S}^{_{(j)}}$ and a controller $\mathcal{C}^{_{(j)}}$. The loops share access to a common medium on the sensor link. A closed-loop system in this network can be modelled as shown in Fig.~\ref{Fig:MAmodel}, with the index $j$ dropped for simplicity. The block $\mathcal{N}$ represents the network as seen by this loop, and the block $\mathcal{R}$ denotes the Contention Resolution Mechanism (CRM), which determines access to the network. Each of the blocks in Fig.~\ref{Fig:MAmodel} is explained below.

\noindent \textbf{Plant: } The plant $\mathcal{P}$ has state dynamics given by
\begin{equation}
\label{Eq:StateSpace4} x_{k+1} = A x_k + B u_k + w_k,
\end{equation}
where $A \in \mathbb{R}^{n \times n}$, $B \in \mathbb{R}^{n \times m}$ and $w_k$ is i.i.d. zero-mean Gaussian
with covariance matrix $R_w$. The initial state $x_0$ is zero-mean Gaussian with covariance matrix $R_0$.

\noindent \textbf{State Based Scheduler: } There is a local scheduler $\mathcal{S}$, situated in the sensor node, between the plant and the controller, which decides if the state is to be sent across the network or not. The scheduler output is denoted $\gamma_k$, where $\gamma_k \in \{0,1\}$. It takes a value $1$ when the state $x_k$ is scheduled to be sent and $0$ otherwise. The scheduling criterion is denoted by the policy $f$, which is defined on the information pattern of the scheduler $\I^{^S}_k$, and is given by
\begin{equation}
\gamma_k = f_k(\vect{u}_{^0}^{_{k-1}}, \omega^s_k) \; , \label{Eq:StateSched4}
\end{equation}
where, $f_k$ is a non-trivial function of $\vect{u}_{^0}^{_{k-1}}$, $\omega^s_k \in \Omega^s_k$, and $\Omega^s_k$ is the sigma-algebra generated by the information set at the scheduler, given by $\I^{^S}_k = \left \{ \vect{x}_{^0}^{_k}, \vect{y}_{^0}^{_{k-1}}, \vect{\gamma}_{^0}^{_{k-1}}, \vect{\delta}_{^0}^{_{k-1}} \right \}$. Here, we use bold font to denote a set of variables such as $\vect{a}_{^t}^{_T} = \{a_t, a_{t+1},\dots,a_T\}$. Note that an explicit acknowledgement (ACK) of a successful transmission is required for $\delta_k$ to be available to the scheduler.

\noindent \textbf{Network: } The network $\mathcal{N}$ generates exogenous traffic, as is indicated by $n_k \in \{0,1\}$. It takes a value $1$ when the network traffic attempts to access the channel, and $0$ otherwise. The network traffic is considered to be stochastic, as it could be generated by another such control loop, or by any other communicating node in the network. Thus, $n_k$ is a binary random variable, which is not required to be i.i.d. We say that there is no exogenous network traffic if $n_k\equiv0$, for all $k$.

\noindent \textbf{CRM: } The CRM block $\mathcal{R}$ resolves contention between multiple simultaneous channel access requests, i.e., when $\gamma_k = 1$ and $n_k = 1$. If the CRM resolves the contention in favour of our control loop, $\delta_k=1$, and otherwise $0$. The CRM can be modelled as the MAC channel response $\mathcal{R}$, with MAC output $\delta_k$ given by
\begin{equation} \label{Eq:delta4}
\delta_k = \mathcal{R}(\gamma_k,n_k)
\end{equation}
For brevity, we also define $\bar{\delta}_k = 1-\delta_k$, which takes a value $1$ when the packet is not transmitted. The MAC channel response $\mathcal{R}$ is modelled as a discrete memoryless channel at the sampling time scale, requiring the CRM to resolve contention with respect to this packet before the next sampling instant. This translates to a limitation on the sampling rates supported by the model.

\noindent \textbf{Measurement: } The measurement across the network is denoted $y_k$. It is a non-linear function of the state $x_k$, and is given by
\begin{equation}
y_k = \delta _k x_k = \begin{cases} x_k & \delta_k=1 \\
0 & \delta_k=0
\end{cases}
\label{Eq:Meas_y_NL4}
\end{equation}
A successful transmission results in the full state being sent to the controller. However, even non-transmissions convey information as the scheduler output $\delta_k$ can be treated as a noisy and coarsely quantized measurement of the state $x_k$.

\noindent \textbf{Controller: } The control law $g$ denotes an admissible policy for the finite horizon $N$ defined on the information pattern of the controller, $\Ick$, and is given by
\begin{equation}
u_k = g_k(\omega^c_k) \; , \label{Eq:Controller4}
\end{equation}
where, $\omega^c_k \in \Omega^c_k$, and $\Omega^c_k$ is the sigma-algebra generated by the information pattern $\Ick = \left \{ \vect{y}_{^0}^{_k}, \vect{\delta}_{^0}^{_k}, \vect{u}_{^0}^{_{k-1}} \right \}$. The objective function, defined over a horizon $N$ is given by
\begin{equation}
\label{Eq:LQGCriterion4} J(f,g)= \E\left[ x_N^T Q_0 x_N + \sum_{s=0}^{N-1} (x_s^T Q_1 x_s + u_s^T Q_2 u_s) \right ]
\end{equation}
where $Q_0$ and $Q_1$ are positive semi-definite weighting matrices and $Q_2$ is positive definite. \newline 

\noindent In the rest of the paper, we address the following questions -
\begin{enumerate}
\item For a NCS with a state-based scheduler, what is the optimal control policy which minimizes the cost $J$ in (\ref{Eq:LQGCriterion4})?
\item Can we find a simple, but sub-optimal, closed-loop system architecture for the given NCS?
\end{enumerate}
To answer the first question, we need to examine whether the system exhibits a dual effect. This also requires us to check if we can find an equivalent system, in the sense of Witsenhausen, for which certainty equivalence holds. The second question requires us to identify restrictions on the scheduling policy $f$, which can ensure separation of the scheduler, controller and observer.

\subsection{Definitions and Properties}

We present a few definitions and properties that are used in the rest of the chapter.

\begin{definition} [Uncontrolled Process] \label{Def:UncontrolledX}
An auxiliary uncontrolled process ($\bar{\mathcal{P}}$) can be defined for any closed-loop system, by removing the effect of the applied control signals from the state. The resulting uncontrolled state is denoted $\bar{x}_k$, and given by
\begin{equation}
\label{Eq:UncontrolledX}
\begin{aligned}
\bar{x}_k &= x_k - \sum_{\ell=1}^{k} A^{\ell-1} B u_{k-\ell} = A^k x_0 + \sum_{\ell=1}^{k} A^{\ell-1} w_{k-\ell} \; .
\end{aligned}
\end{equation}
\end{definition}

\begin{remark} [Last Received Packet Index:]
The time index of the last received packet is denoted $\tau_k$ at time $k$ (illustrated in Fig.~\ref{Fig:TimeLines}), and is given by
\begin{equation} \label{Eq:tau4}
\tau_k = \max\{t: \delta_t=1, \text{ for } -1 \le t \le k \} \text{ and } \delta_{-1}=1, -1 \le \tau_k \le k
\end{equation}

An iterative relationship for $\tau_k$ can be found as
\begin{equation} \label{Eq:tau_Iter4}
\tau_k = \bar{\delta}_k \tau_{k-1} + \delta_k k, \quad \tau_{-1} = -1
\end{equation}
If a packet arrives at current time $k$, the last received packet index $\tau_k = k$. But, if there is no packet at time $k$, then the last received packet index is the same as the last received packet index from time $k-1$, i.e., $\tau_k = \tau_{k-1}$. This implies that $\tau_k \in \{-1,\dots,k\}$. \newline
\end{remark}

\begin{remark} [Dual Effect: ] Note that the control $u_k$ might affect the future state uncertainty, in addition to its direct effect on the state. This is called the dual effect of control \cite{Feldbaum1961}, and is discussed for state-based schedulers in Section~\ref{SS:DualEffect}.
\begin{definition} [No dual effect \cite{Bar-Shalom1974}]
A control signal is said to have no dual effect of order $r\ge2$, if
\begin{equation}
\E[M^r_{k,i}|\Ick] = \E[M^r_{k,i}|x_0,\vect{w}_{^0}^{_{\tau_k}},\vect{n}_{^0}^{_k}] \label{Eq:Def_DualEffect4}
\end{equation}
where $M^r_{k,i} = ( x_{k,i}-\E[x_{k,i}|\Ick] )^r$ is the $r^\textrm{th}$ central moment of the $i^\textrm{th}$ component of the state $x_{k,i}$ conditioned on $\Ick$ and $\tau_k$ is the time index of the last received measurement at time $k$.
\end{definition}
Note that $M^r_{k,i}$ in (\ref{Eq:Def_DualEffect4}) must specifically not be a function of the past control policies
$\vect{g}_{^{0}}^{_{k-1}}$ for the control signal to have no dual effect of order $r$. In other words, if there is no dual effect,
the expected future uncertainty is not affected by the controls $\vect{u}_{^0}^{_{k-1}}$. In the presence of a dual effect, the optimal control laws are hard to find \cite{AAstrom1995}. \newline
\end{remark}

\begin{remark} [Certainty Equivalence: ]
There are two closely related terms: a certainty equivalent controller and the Certainty Equivalence Principle. We define both these terms with respect to the deterministic optimal controller, with full state information, for the above problem setup \cite{AAstrom1970,Bar-Shalom1974}. These properties are discussed for state-based schedulers in Section~\ref{SS:CE}.

\begin{definition} [Certainty Equivalent Controller]
A certainty equivalent controller\index{Certainty Equivalence Principle!certainty equivalent control} uses the deterministic optimal controller, with the state $x_k$ replaced by the estimate $\hat{x}_{^{k|k}} = \E[x_k|\Ick]$, as an ad hoc control procedure.
\end{definition}

Sometimes, there is no loss in optimality in using a certainty equivalent controller. Then, we say that the Certainty Equivalence Principle holds.
\begin{definition} [Certainty Equivalence Principle]
The Certainty Equivalence Principle holds if the closed-loop optimal controller has the same form as the deterministic optimal controller with the state $x_k$ replaced by the estimate $\hat{x}_{^{k|k}}$. \newline
\end{definition}
\end{remark}

\begin{remark} [Correlated Network Noise: ] We state a property of feedback systems with a state-based scheduler which share a contention-based multiple access network. Even if the initial states and disturbances of all the plants in the network are independent, the contention-based MAC introduces a correlation between the traffic sources in the network, as noted in \cite{Cervin2008,Rabi2009}.
\begin{lemma}
\label{L:CorrSys}
For a closed-loop system defined by (\ref{Eq:StateSpace4})--(\ref{Eq:Controller4}), the exogenous network traffic indicated by $n_k$ is correlated to the state of the plant $x_k$.
\end{lemma}

\begin{IEEEproof}
The MAC output $\delta_{k-1}$ is a function of the state $x_{k-1}$ and the indicator of network traffic $n_{k-1}$, from (\ref{Eq:StateSched4}) and (\ref{Eq:delta4}). The control signal $u_{k-1}$ is a function of the MAC output $\delta_{k-1}$ from (\ref{Eq:Controller4}), and is applied through feedback to the plant. Thus, $x_{k}$ and $\gamma_k$ are correlated to $\delta_{k-1}$. Similarly, the network traffic from other closed-loop systems (and its indicator $n_k$) is correlated to $\delta_{k-1}$, and consequently, $x_k$.
\end{IEEEproof}
\end{remark}

%%%%%%%%%%%%%%%%%%%%%%%%%%%%%%%%%%%%%%%%%%%%%%%%%%%%%%%%%%%%%%%%%%%%%%%%%%%%%%%%

\section{Optimal Controller Design} \label{S:SSched}

\IEEEPARstart{W}{e} present the main results of this paper in this section. We first analyze the effects of a state-based scheduler on a control loop with no exogenous network traffic, i.e., $n_k = 0$. As a consequence of this, the MAC output is equal to the scheduler output, i.e., $\delta_k = \gamma_k$. We show that there is a dual effect of the control signal, and that the scheduling policy must be restricted from using the past control inputs for the Certainty Equivalence Principle to hold. We illustrate this for a second order system with a state-based scheduler in Fig.~\ref{Fig:DualRole}, and show that the controller is not oblivious to the scheduler boundaries. We extend our results to the case with exogenous network traffic.

\subsection{Dual Effect with State-based Scheduling} \label{SS:DualEffect}

\begin{figure}[!t]
\begin{center}
\subfigure[Delay and Index of Last Received Packet]{\label{Fig:TimeLines}
\includegraphics[scale=0.5]{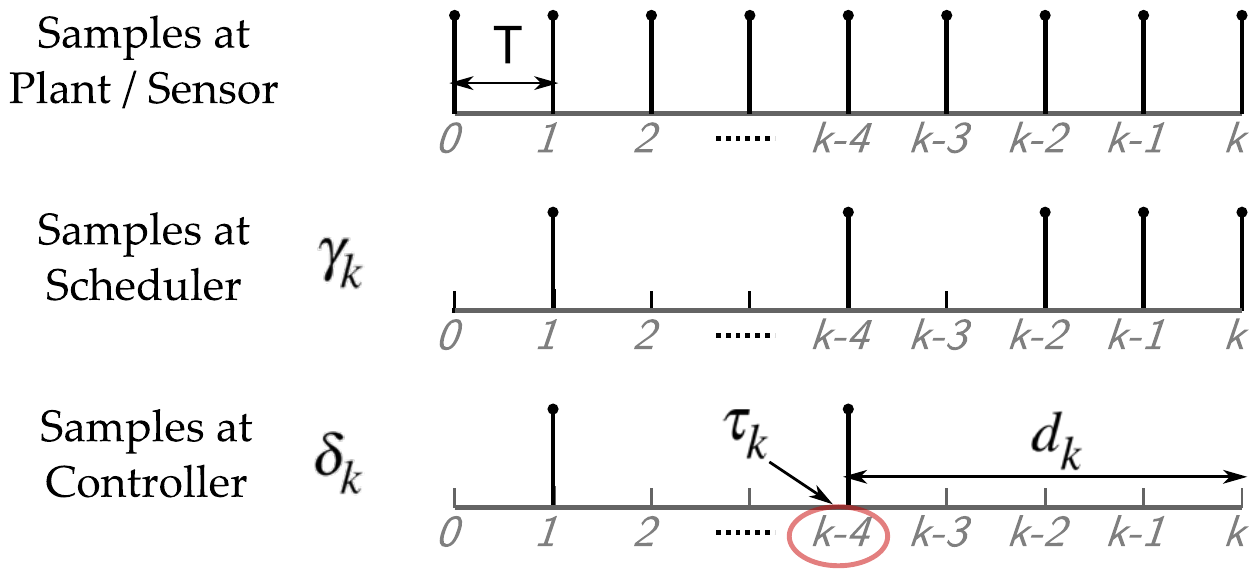}} \hspace{2cm}
\subfigure[Dual Control Incentives]{\label{Fig:DualRole}
\includegraphics[scale=1]{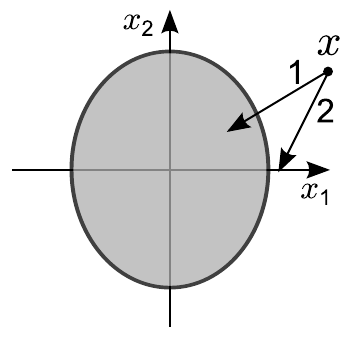}}
\caption{An illustration of the delay since the last received packet $(d_k)$ and the index of the last received packet $(\tau_k)$ in (a). In (b), the shaded area represents the non-scheduled region for a state $x\in\mathbb{R}^2$, as defined by the scheduling policy $f$. States in this region are not considered to be events and correspondingly result in a non-zero estimation error. Outside this region, the estimation error is zero. Thus, the controller has an incentive to move some states along path $2$, as compared to path $1$.}
\end{center}
\vspace{-5mm}
\end{figure}

For the problem defined in Section~\ref{S:ProbSetup4}, we observe the following result.

\begin{theorem} \label{T:DualEffect_nNW}
For the closed-loop system defined by (\ref{Eq:StateSpace4})--(\ref{Eq:Controller4}), with no exogenous network traffic, the control signal has a dual effect of order $r=2$.
\end{theorem}

\begin{IEEEproof}
We examine the estimation error, and show that it is not equivalent to the estimation error generated by the uncontrolled process ($\bar{\mathcal{P}}$ from Definition~\ref{Def:UncontrolledX}) in the place of the plant $\mathcal{P}$. Thus, we proove that the estimation error covariance is a function of the applied controls $\vect{u}_{^0}^{_{k-1}}$.

From (\ref{Eq:Meas_y_NL4}), we know that a successful transmission results in the full state being sent to the controller, whereas a non-transmission conveys only a single-bit of information ($\delta_k$ is binary) about the state to the controller. Thus, the estimate, $\hat{x}_{^{k|k}} \triangleq \E[x_k|\Ick]$, is given by
\begin{equation*}
\hat{x}_{^{k|k}} = \delta_k x_k + \bar{\delta}_k \E[ x_k | \Ick,\delta_k=0 ] \; .
\end{equation*}
This estimate always depends on $\delta_k$, due to the asymmetry in the resolution of the received information with and without a transmission. The scheduler outcome, and consequently $\delta_k$, are influenced by the applied control inputs $\vect{u}_{^0}^{_{k-1}}$ in a state-based scheduler such as (\ref{Eq:StateSched4}). The estimation error, defined as $\tilde{x}_{^{k|k}} \triangleq x_k - \E[x_k|\Ick]$, is given by
\begin{equation} \label{Eq:DE_EstError}
\tilde{x}_{^{k|k}} = (x_k-\E[x_k|\Ick,\delta_k = 0]) \cdot \bar{\delta}_k \; ,
\end{equation}
and can also be seen to depend on $\delta_k$. The estimation error when there is no transmission is defined as $\tilde{x}_{^{k|k}}^{_{0}} \triangleq x_k-\E[x_k|\Ick,\delta_k = 0]$, and is given by
\begin{align*}
\tilde{x}_{^{k|k}}^{_{0}} &= A^k x_0 + \sum_{\ell=1}^{k} A^{\ell-1} (B u_{k-\ell} + w_{k-\ell}) - \E[ A^k x_0 + \sum_{\ell=1}^{k} A^{\ell-1} (B u_{k-\ell} + w_{k-\ell}) | \Ick, \delta_k=0] \\
&= \bar{x}_k - \E[ \bar{x}_k | \Ick, \delta_k=0] \; ,
\end{align*}
where $\bar{x}_k$ is the state of the uncontrolled process (see Definition~\ref{Def:UncontrolledX}). Thus, the first term of the product in (\ref{Eq:DE_EstError}) remains unchanged if the plant is replaced by the uncontrolled process. However, the effect of the applied controls cannot be removed from $\bar{\delta}_k = (1-\delta_k)$, which is the second term of the product in (\ref{Eq:DE_EstError}).

Thus, the estimation error is always dependent on the applied controls and this distinguishes the current problem from other related problems, such as in \cite{Tatikonda2004,Nair2007}.
The error covariance, $P_{^{k|k}} \triangleq \E[\tilde{x}_{^{k|k}}\tilde{x}_{^{k|k}}^T | \Ick]$, is given by
\begin{equation}
\label{Eq:DualP}
P_{^{k|k}} = \bar{\delta}_k \cdot (\E[ \tilde{x}_{^{k|k}}\tilde{x}_{^{k|k}}^T |\Ick,\delta_k = 0]) \; .
\end{equation}
The covariance $P_{^{k|k}}$ is zero if the scheduling criterion in (\ref{Eq:StateSched4}) is fulfilled, and non-zero otherwise. Clearly, $P_{^{k|k}}$ is a function of the past controls. Hence, $P_{^{k|k}}$ does not satisfy the condition (\ref{Eq:Def_DualEffect4}) required to have no dual effect. Thus, we see that the system (\ref{Eq:StateSpace4})--(\ref{Eq:Controller4}) exhibits a dual effect of order $r=2$.
\end{IEEEproof}

In this setup, there is an incentive for the control policy to modify the estimation error along with controlling the plant, as illustrated in Fig.~\ref{Fig:DualRole}. Thus, the controller might prefer to move some states along path $2$, as compared to path $1$, to improve the estimation error.

\subsection{Equivalent Schedulers vs. Equivalent Systems}
\label{SSS:EquiClass}

Every state-based scheduler $f$, defined in (\ref{Eq:StateSched4}), can be transformed into an equivalent scheduler $\tilde{f}$, such as
\begin{equation}
\gamma_k = \tilde{f}_k(\tilde{\omega}^s_k) \; ,
\label{Eq:TildeSched}
\end{equation}
where, $\tilde{\omega}^s_k \in \tilde{\Omega}^s_k$, and $\tilde{\Omega}^s_k$ is the sigma-algebra generated by the information pattern $\tilde{\I}^{^S}_k = \{ x_0, \vect{w}_{^0}^{_{k-1}} \}$. The applied controls $\vect{u}_{^0}^{_{k-1}}$ can be reconstructed at the scheduler at time $k$, and hence, such a transformation can always be accomplished. We now examine the question of whether the closed-loop system with this equivalent scheduler, is equivalent to the original system. Witsenhausen \cite{Witsenhausen1971} defines an equivalent design, which gives us the following definition when applied to our problem.

\begin{definition} \label{Def:EquiSys}
An equivalent (in the sense of Witsenhausen) control design $g_{\textrm{eq}}$ for the optimal controller $g^{_*}$, which minimizes the cost criterion (\ref{Eq:LQGCriterion4}) for the system defined by (\ref{Eq:StateSpace4})--(\ref{Eq:Controller4}), satisfies the equivalence relationship given by
\begin{equation}
\vect{u}^{_*} = \Upsilon(\vect{\omega}, g^{_*}) = \Upsilon(\vect{\omega}, g_{^\textrm{eq}}) \; ,
\label{Eq:EquiRelation_g}
\end{equation}
where $\Upsilon$ is obtained by recursive substitution for the control signals in the system equations with the respective control policy and the primitive random variables $\vect{\omega}_{k} = [x_0, \vect{w}_{^{0}}^{_{k-1}}]$.
\end{definition}

For brevity, we adopt the following notation. Let $\{\mathcal{P},f_1,g_1\}$ denote a system with the plant given by (\ref{Eq:StateSpace4}), with $f_1$ as the given scheduler and $g_1$ as the optimal controller for the cost criterion (\ref{Eq:LQGCriterion4}) applied to this system. We now note the following result.
\begin{theorem} \label{T:NEquiSys}
For two schedulers $f$, given by (\ref{Eq:StateSched4}), and $\tilde{f}$, given by (\ref{Eq:TildeSched}), which result in the same schedules for the closed-loop system given by (\ref{Eq:StateSpace4}), (\ref{Eq:delta4})--(\ref{Eq:Controller4}), with no exogenous network traffic, $\{\mathcal{P},\tilde{f},\tilde{g}\}$ is not an equivalent system to $\{\mathcal{P},f,g^{_*}\}$, in the sense of Witsenhausen.
\end{theorem}

\begin{IEEEproof}
Definition \ref{Def:EquiSys} requires the control signals obtained using the policies $g^{_*}$ and $\tilde{g}$ to be equal. In this proof, we find the optimal control policies for $\{\mathcal{P},\tilde{f},\tilde{g}\}$ and $\{\mathcal{P},f,g^{_*}\}$, and show that they do not result in the same control signals.

For the optimal control policy, which minimizes the quadratic cost $J$ in (\ref{Eq:LQGCriterion4}), to be certainty equivalent, we need to find a solution to the Bellman equation \cite{AAstrom1970}, which is a one-step minimization of the form
\begin{equation} \label{Eq:Bellman4}
V_k = \min _{u_k} \; \E[x_k^T Q_1 x_k + u_k^T Q_2 u_k + V_{k+1}|\Ick] \; .
\end{equation}
In general, without defining a structure for the estimator, the solution to the functional is given \cite{Bar-Shalom1974} in the form of
\begin{equation}
V_k = \E\left[x_k^T S_k x_k | \Ick \right] + s_k \; ,
\label{Eq:FuncSol4}
\end{equation}
where $S_k$ is a positive semi-definite matrix and both $S_k$ and $s_k$ are not functions of the applied control signals $\vect{u}_{^0}^{_{k-1}}$. We now prove that a solution of this form can be found for $\{\mathcal{P},\tilde{f},\tilde{g}\}$, but not for $\{\mathcal{P},f,g^{_*}\}$.

First consider the system $\{\mathcal{P},\tilde{f},\tilde{g}\}$. We denote the state and control signals of this system as $\tilde{x}_k$ and $\tilde{u}_k$. At time $N$, the functional has a trivial solution with $S_N=Q_0$ and $s_N=0$. This solution can be propagated backwards, in the absence of a dual effect. To show this, we use the principle of induction, and assume that a solution of the form (\ref{Eq:FuncSol4}) holds at time $k+1$. Then, at time $k$, we have
\begin{equation}
\label{Eq:SimplifyV}
\begin{aligned}
V_k &= \min _{u_k} \; \E[\tilde{x}_k^T Q_1 \tilde{x}_k + \tilde{u}_k^T Q_2 \tilde{u}_k + \tilde{x}_{k+1}^T S_{k+1} \tilde{x}_{k+1}  + s_{k+1}| \Ick] \notag \\
&= \min _{u_k} \; \E[\tilde{x}_k^T (Q_1 + A^T S_{k+1} A) \tilde{x}_k | \Ick] + \tr\{S_{k+1} R_w\} + \E[s_{k+1}|\Ick] \notag \\
& \quad \quad \; + \tilde{u}_k^T (Q_2 + B^T S_{k+1} B)\tilde{u}_k + \hat{\tilde{x}}_{^{k|k}}^T A^T S_{k+1} B \tilde{u}_k + \tilde{u}_k^T B^T S_{k+1} A \hat{\tilde{x}}_{^{k|k}} \; ,
\end{aligned}
\end{equation}
where $\hat{\tilde{x}}_{^{k|k}} \triangleq \E[\tilde{x}_k | \Ick]$. The optimal control is found to be
\begin{equation}
\label{Eq:uCE}
\begin{aligned}
\tilde{u}_k &= - L_k \hat{\tilde{x}}_{^{k|k}} \; , \; L_k = (Q_2 + B^T S_{k+1} B)^{-1} B^T S_{k+1} A \; .
\end{aligned}
\end{equation}
Substituting the expression for $\tilde{u}_k$ into $V_k$ gives us a solution of the form in (\ref{Eq:FuncSol4}), with
\begin{align}
S_k ={}& Q_1 + A^T S_{k+1} A - A^T S_{k+1} B (Q_2 + B^T S_{k+1}B)^{-1} B^T S_{k+1} A \; , \notag \\
s_k ={}& \tr\{S_{k+1} R_w\} + \E[s_{k+1}|\Ick] + \tr\{A^T S_{k+1} B (Q_2 + B^T S_{k+1}B)^{-1} B^T S_{k+1} A P_{^{k|k}}\} \; , \label{Eq:Iter_s_k}
\end{align}
where the matrix $S_k$ is positive semi-definite and not a function of the applied controls $\vect{\tilde{u}}_{^0}^{_{k-1}}$. The scalar $s_k$ is not a function of the applied controls $\vect{\tilde{u}}_{^0}^{_{k-1}}$ if and only if $P_{^{k|k}}$ has no dual effect \cite{Bar-Shalom1974}. From the expression for the error covariance $P_{^{k|k}}$ (\ref{Eq:DualP}), it is clear that a scheduling criterion that is not a function of the past control actions, such as (\ref{Eq:TildeSched}), results in no dual effect. Under this condition, $s_k$ is not a function of the applied controls $\vect{\tilde{u}}_{^0}^{_{k-1}}$ and the proof by induction is complete. Since the optimal control signal (\ref{Eq:uCE}) is a function of only the estimate $\hat{\tilde{x}}_{^{k|k}}$, the Certainty Equivalence Principle holds.

Now, consider the system $\{\mathcal{P},f,g^{_*}\}$ with state $x_k$ and control $u^*_k$. Solving the backward recursion as we did above, we find that $V_N$ and $V_{N-1}$ have a solution of the form (\ref{Eq:FuncSol4}), with $S_N = Q_0$ and $s_N = 0$, and $S_{N-1}$ and $s_{N-1}$ given by (\ref{Eq:Iter_s_k}) with $k=N-1$. However, $V_{N-2}$ results in a different minimization problem for this system because of the dual effect in $\{\mathcal{P},f,g^{_*}\}$, as indicated next. The optimal control signal $u^{_*}_{N-2}$ can be obtained by solving
\begin{align*}
\frac{\partial V_{N-2}}{\partial u^{_*}_{N-2}} ={}& 2 u^{{_*} T}_{N-2} (Q_2+B^T S_{N-1} B) + 2 \hat{x}^T_{^{N-2|N-2}} A^T S_{N-1} B \\
&+ \frac{\partial}{\partial u^{_*}_{N-2}} \bigg( \tr\{ A^T S_N B (Q_2+B^T S_N B)^{-1} B^T S_N A \cdot \E[P_{^{N-1|N-1}}|\I^{^C}_{N-2}] \} \bigg) = 0 \; .
\end{align*}
Multiplying the above expression with $(Q_2+B^T S_{N-1} B)^{-1}$ from the right and using (\ref{Eq:uCE}) to denote $u^{CE}_{N-2} = -L_{N-2} \hat{x}_{^{N-2|N-2}}$, we obtain the simpler equation
\begin{equation}
\frac{\partial V_{N-2}}{\partial u^{_*}_{N-2}} = 2 (u^{{_*} T}_{N-2} - u_{N-2}^{CE,T}) + \frac{\partial}{\partial u^{_*}_{N-2}} \bigg( \tr\{ \mathds{K}_{N-2} \E[P_{^{N-1|N-1}}|\I^{^C}_{N-2}] \} \bigg) = 0 \; ,
\label{Eq:minV_N-2_DE}
\end{equation}
where, we set $\mathds{K}_{N-2} = (Q_2+B^T S_{N-1} B)^{-1} A^T S_N B (Q_2+B^T S_N B)^{-1} B^T S_N A $. The last term in (\ref{Eq:minV_N-2_DE}), related to the estimation error covariance $P_{^{N-1|N-1}}$, is not equal to zero as implied by the dual effect property from Theorem~\ref{T:DualEffect_nNW}. Due to this term, the above minimization problem is not linear, and thus, the solutions $u^{CE}_{N-2}$ and $u^*_{N-2}$ are not equal. Since $u^{CE}_{N-2}$ has the same form as $\tilde{u}_{N-2}$, we also note that $\tilde{u}_{N-2}$ and $u^*_{N-2}$ have very different forms. From this point on, the cost-to-go for the optimal control policy $g^{_*}$ does not have a solution of the form given by (\ref{Eq:FuncSol4}). Hence, the control signals $\{\vect{\tilde{u}}\}_{^0}^{_{N-3}}$ and $\{\vect{u}^{_*}\}_{^0}^{_{N-3}}$ will not be equal. Now, the joint distribution of all system variables could be quite different for schedulers $\tilde{f}$ and $f$. Thus, the described transformation of the scheduling criterion does not result in an equivalent class construction.
\end{IEEEproof}

Due to the dual effect, the optimal control action takes on two roles. One, to control the plant, and the other, to probe the plant state which could result in an improved estimate \cite{AAstrom1995}. In the certainty equivalent setup, the probing action cannot be implemented due to the lack of a dual effect and the resulting control actions will not remain the same.

\subsection{Conditions for Certainty Equivalence} \label{SS:CE}
From the previous discussions, it is clear that a scheduling criterion independent of the past control actions results in no dual effect. This result is presented below.

\begin{corollary} \label{C:CE}
For the closed-loop system defined by (\ref{Eq:StateSpace4})--(\ref{Eq:Controller4}), with no exogenous network traffic, the optimal controller, with respect to the cost in (\ref{Eq:LQGCriterion4}), is certainty equivalent if and only if the scheduling decisions are not a function of the applied control actions, such as in (\ref{Eq:TildeSched}).
\end{corollary}

\begin{IEEEproof}
In the proof of Theorem~\ref{T:NEquiSys}, it is clear from (\ref{Eq:uCE}) that the optimal control policy $\tilde{g}$ for the system  $\{\mathcal{P},\tilde{f},\tilde{g}\}$ is certainty equivalent.

To show the necessity of this condition for certainty equivalence, we need to show that if the optimal control signal has the form in (\ref{Eq:uCE}) at time $k$, then the scheduling policy is not a function of the controls for $n<k$, for all $k$. Accordingly, assume that the optimal control signal is given by (\ref{Eq:uCE}) for $k=N-1,\dots,n+1$. Then, the optimal cost-to-go is of the form in (\ref{Eq:FuncSol4}), at time $n+1$ and
\begin{equation*}
s_{n+1} = \sum_{k=n+1}^{N-1} \E[\tr\{A^T S_{k+1} B (Q_2 + B^T S_{k+1}B)^{-1} B^T S_{k+1} A P_{^{k|k}} + S_{k+1} R_w\} | \Ick] \; , 
\end{equation*}
when written out explicitly. We know that the optimal control signal $u_n$ is obtained by minimizing (\ref{Eq:Bellman4}) at time $n$. This control signal will have the form in (\ref{Eq:uCE}) for all $Q_2>0$ only if $s_{n+1}$ is independent of $u_n$, or if the estimation error covariances $P_{^{k|k}}$, for $k=\{n+1,\dots,N-1\}$, are not a function of $u_n$. From the result in Theorem~\ref{T:DualEffect_nNW}, this is only possible when the scheduling policy is not a function of $u_n$. Since this is true for $n=0,\dots,N-1$, the scheduling policy must not be a function of $\vect{u}_{^{0}}^{_{k-1}}$.
\end{IEEEproof}

Corollary \ref{C:CE} provides us with a restriction on the scheduler to guarantee certainty equivalence. Note that the resulting closed-loop system is not equivalent to the original problem setup, as shown in Theorem~\ref{T:NEquiSys}.

\subsection{Effect of State-based Schedulers with exogenous network traffic} \label{SS:withNetwork}
In this subsection, we re-analyze the effects of a state-based scheduler on the control loop in the \emph{\textbf{presence}} of exogenous network traffic. Thus, we have $n_k \neq 0$ and a channel output given by (\ref{Eq:delta4}). Recall from Lemma~\ref{L:CorrSys}, that the network traffic indicator $n_k$ is correlated to the state of the plant $x_k$. The Certainty Equivalence Principle need not hold for plants where the measurement noise is correlated to the process noise \cite{Bar-Shalom1974}. To focus on the effect of state-based schedulers on the closed-loop system, the results presented in the previous subsection did not include exogenous network traffic. Now, we re-derive some of the above results for the system in the presence of exogenous network traffic. \newline

\begin{lemma} \label{L:DualEffect_withNetwork}
For the closed-loop system defined by (\ref{Eq:StateSpace4})--(\ref{Eq:Controller4}), the control signal has a dual effect of order $r=2$.
\end{lemma}

\begin{IEEEproof}
The MAC output $\delta_k$ (\ref{Eq:delta4}) is clearly still a function of the applied controls, through the state-based scheduler outcome. Thus, the estimation error covariance $P_{^{k|k}}$, in (\ref{Eq:DualP}), remains a function of the applied controls $\vect{u}_{^0}^{_{k-1}}$. Since $P_{^{k|k}}$ does not satisfy the condition (\ref{Eq:Def_DualEffect4}) required to have no dual effect, we see that the system (\ref{Eq:StateSpace4})--(\ref{Eq:Controller4}) exhibits a dual effect of order $r=2$.
\end{IEEEproof}

With the above result, Theorem~\ref{T:NEquiSys} can be easily extended to include the case with exogenous network traffic. However, it is not as straightforward to extend Corollary~\ref{C:CE}. When the measurement noise is correlated to the process noise, certainty equivalence need not hold. To see why, recall the proof of Theorem~\ref{T:NEquiSys}, where we derive a solution of the form $V_k = \E[x_k^T S_k x_k | \Ick ] + s_k$ for the Bellman equation (\ref{Eq:Bellman4}). Now, if $w_k$ is correlated to the variables in the information set $\Ick$, specifically $\vect{n}_{^0}^{_k}$, the minimization with respect to $u_k$ in (\ref{Eq:SimplifyV}) must include the term $\tr\{S_{k+1} R_w\}$. Then, the optimal controller will not have the form shown in (\ref{Eq:uCE}), and certainty equivalence will not hold.

We need to prove that $w_k$ is independent of $\vect{n}_{^0}^{_k}$ for the Certainty Equivalence Principle to hold, which we do below.

\begin{corollary} \label{C:CEnetwork}
For the closed-loop system defined by (\ref{Eq:StateSpace4})--(\ref{Eq:Controller4}), the optimal controller, with respect to the cost criterion (\ref{Eq:LQGCriterion4}), is certainty equivalent if the exogenous network traffic indicator $n_k$ is independent of the process noise $w_k$, and, if the scheduling decisions are not a function of the applied controls, i.e., if
\begin{equation}
\gamma_k = \check{f}_k(\check{\omega}^s_k) \; ,
\label{Eq:CheckSched}
\end{equation}
where, $\check{\omega}^s_k \in \check{\Omega}^s_k$, and $\check{\Omega}^s_k$ is the sigma-algebra generated by the information set $\check{\I}^{^S}_k = \{ x_0,\vect{w}_{^0}^{_{k-1}},\vect{n}_{^0}^{_{k-1}} \}$.
\end{corollary}

\begin{IEEEproof}
Note that $n_k$ is only correlated to $\vect{\delta}_{^0}^{_k}$ and thus, to the signals $\vect{w}_{^0}^{_{k-1}}$, from Lemma~\ref{L:CorrSys}. As the process noise is i.i.d, $n_k$ is independent with respect to $w_k$. A scheduler of the form (\ref{Eq:CheckSched}) is not a function of the applied controls, and thus, Certainty Equivalence holds.
\end{IEEEproof}

\section{Closed-Loop System Architecture} \label{S:CSArch}
\IEEEPARstart{I}{n} this section, we identify a property of the scheduling policy that results in a simplification of the design of the closed-loop system. This enables us to propose a dual predictor architecture for the closed-loop system, which results in a separation of the scheduler, observer and controller designs.

\subsection{Observer Design}
In this section, we propose a structure for the estimator at the controller. Due to the non-linearity of the problem, the estimate in general can be hard to compute.

The estimation error is reset to zero with every transmission, as we send the full state. Consider one such reset instance, a time $k$ such that $\delta_k=1$. The state is sent across the network, $y_k=x_k$, so the estimate $\hat{x}_{^{k|k}} = x_k$. A suitable control signal $u_k$ is
found and applied to the plant, which results in the next state $x_{k+1}$. Now, the scheduler can generate one of two outcomes. We consider each case, and find an expression for the estimate, below:
\newcounter{A1counter}
\begin{list}{\alph{A1counter}) }
{\usecounter{A1counter}} \item $\delta_{k+1}=0$: We need an estimate of $w_k$. We use the scheduler output as a coarse quantized measurement to generate this, as follows:
\begin{align}
\hat{x}_{^{k+1|k+1}} &= \E[x_{k+1}|\I^{^C}_{k+1}, \delta_{k+1}=0] \notag \\
&= Ax_k + Bu_k + \E[w_{k}|\grave{f}(w_k) = 0] \; ,\label{Eq:GraveF}\\
\tilde{x}_{^{k+1|k+1}} &\triangleq x_{k+1}-\hat{x}_{^{k+1|k+1}} = w_k - \E[w_{k}|\grave{f}(w_k) = 0] \; , \notag
\end{align}
where, $\grave{f}(w_k) \equiv f(Ax_k+Bu_k+w_k | x_k,u_k)$.

\item $\delta_{k+1}=1$: The estimation error is zero as $\hat{x}_{^{k+1|k+1}} = x_{k+1}$.
\end{list}
The transformation to $\grave{f}$ in (\ref{Eq:GraveF}), is not intended to remove the dual effect, but merely serves to remove the known variables from the expression. The dual effect has influenced the packet's transmission, i.e., the value of $\delta_{k+1}$.

To see this more clearly, we look at the next time instant. Now a signal $u_{k+1}$ is generated, and applied to the plant. We note that $x_{k+2} = A^2 x_k + ABu_k + Bu_{k+1} + Aw_k + w_{k+1}$. The state $x_{k+2}$ is either sent to the controller or not depending on the scheduler outcome $\delta_{k+2}$. Again, we look at both cases, and derive an expression for the estimate:
\newcounter{A2counter}
\begin{list}{\roman{A2counter}) }
{\usecounter{A2counter}} \item $\delta_{k+2}=0$: We now need to estimate $Aw_k + w_{k+1}$, as the rest is completely known from $x_{k+2}$. We use both scheduler outputs $\delta_{k+1}$ and $\delta_{k+2}$ to generate an estimate of the unknown variables as
\begin{align*}
\hat{x}_{^{k+2|k+2}} ={}& A^2x_k + ABu_k + Bu_{k+1} + \E[Aw_k+w_{k+1}|\grave{f}(w_k)=0, \grave{f}(Aw_k+w_{k+1}) = 0] \; ,\\
\tilde{x}_{^{k+2|k+2}} ={}& Aw_k+w_{k+1} - \E[Aw_k+w_{k+1}|\grave{f}(w_k)=0, \grave{f}(Aw_k+w_{k+1}) = 0] \; .
\end{align*}
\item $\delta_{k+2}=1$: The estimation error is zero as $\hat{x}_{^{k+2|k+2}} = x_{k+2}$.
\end{list}
This process can be continued recursively through a non-transmission burst, until finally a measurement is received and the estimation error is reset to zero. Thus, the observer computes the estimate at any time $k$ as
\begin{equation} \label{Eq:Estimate}
\hat{x}_{^{k|k}} =
\begin{cases}
x_k, & \delta_k=1, \\
\begin{aligned}
&A^{k-\tau_k} x_{\tau_k} + \sum_{s=1}^{k-\tau_k} A^{s-1} B u_{k-s} + \E[\sum_{s=1}^{k-\tau_k} A^{s-1} w_{k-s}|\grave{f}_k,..,\grave{f}_{\tau_k+1} = 0],
\end{aligned} & \delta_k=0,
\end{cases}
\end{equation}
where $\tau_k$ is the time index of the last received measurement at time $k$, as defined in (\ref{Eq:tau4}), and the argument to the function $\grave{f}_t$ is given by the term $\sum_{s=1}^{t-\tau_t} A^{s-1} w_{t-s}$.

\subsection{State-based Scheduler Design: Symmetric Schedulers}
The computation of the term $\E[\sum_{s=1}^{k-\tau_k} A^{s-1} w_{k-s}| \grave{f}_k,.., \grave{f}_{\tau_k+1} = 0]$, for a burst of non-transmissions of length greater than one, makes the estimate given in (\ref{Eq:Estimate}) hard to evaluate. This is because the quantized noise is not Gaussian. As a sub-optimal, but simplified approach, consider the scheduling criterion given by any symmetric map $f^{sym}(r) = f^{sym}(-r)$ with
\begin{equation}
\label{Eq:Symf}
\begin{aligned}
\gamma_k ={}& f^{sym} (\sum_{s=1}^{k-\tau_{k-1}} A^{s-1} w_{k-s}) \; .
\end{aligned}
\end{equation}
Since $\tau_k$ is not defined without the MAC output $\delta_k$ in (\ref{Eq:tau4}), we replace it with $\tau_{k-1}$, which is also a measure of the non-transmission burst. Choosing the scheduler in this manner results in a zero mean estimate from the quantized noise when there is no transmission. Now, the estimate is easy to compute and the observer can be designed without knowledge of the scheduling policy. Also, a certainty equivalent control can be applied. This observation is summarized below, and is used to design the scheduler presented in Section~\ref{SS:DualPred}.

\begin{proposition} \label{P:OptObserver}
For the closed-loop system defined by (\ref{Eq:StateSpace4})--(\ref{Eq:LQGCriterion4}), the use of the symmetric scheduling policy defined in (\ref{Eq:Symf}) implies that certainty equivalence holds, and it also results in separation in design, between the estimator and scheduler.
\end{proposition}

\subsection{The Dual Predictor Architecture} \label{SS:DualPred}

In this section, we examine closed-loop design of the complete system, including scheduler, observer and controller. From the results of Lemma~\ref{L:DualEffect_withNetwork} and Proposition \ref{P:OptObserver}, it is clear that the scheduler, observer and controller designs are coupled. It is not possible to design the optimal scheduling policy independently and combine it with a certainty equivalent controller and optimal observer to get the overall optimal closed-loop system. At the same time, solving for the jointly optimal scheduler, observer and controller is a hard problem.

\begin{figure}[tpb]
\begin{center}
\includegraphics[scale=0.65]{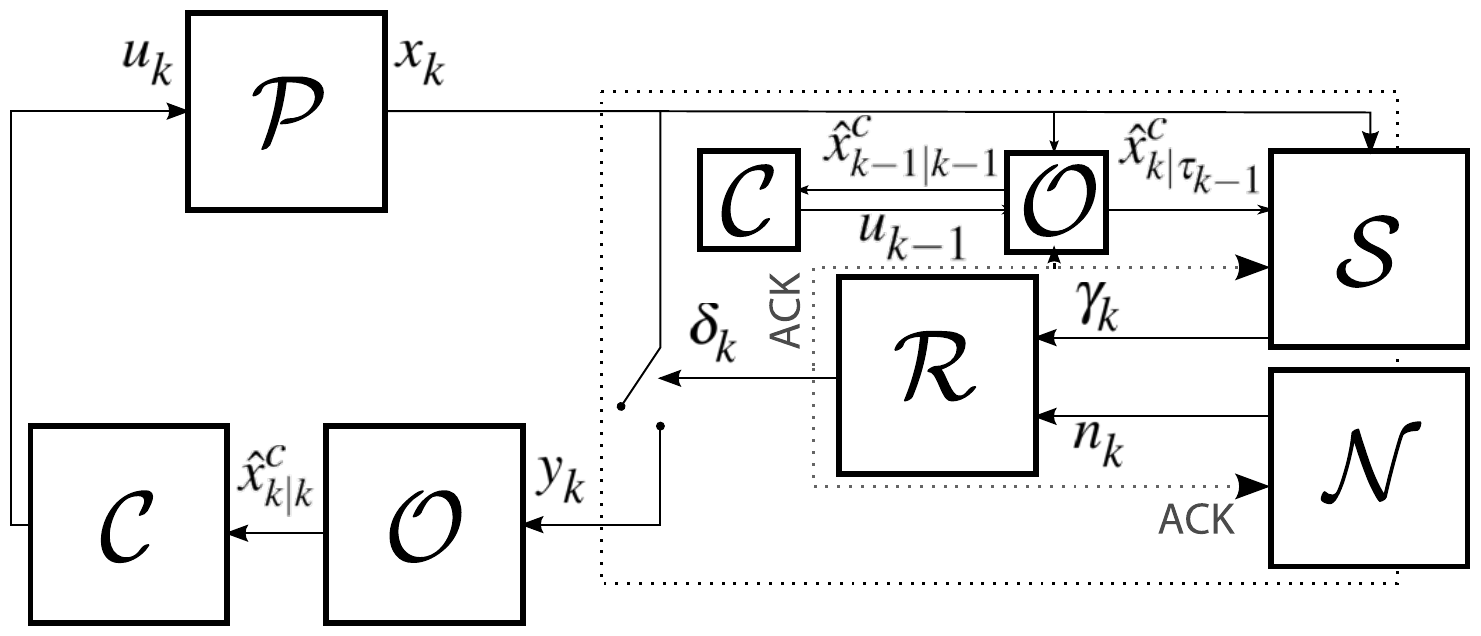}
\caption[The Dual Predictor Architecture]{State-based Dual Predictor Architecture: the innovations to the observer serve as input to the scheduler. The resulting setup is certainty equivalent. The observer is simple, and computes the MMSE estimate.} \label{Fig:StateDualPred}
\vspace{-5mm}
\end{center}
\end{figure}

Thus, we propose an architecture, shown in Fig. \ref{Fig:StateDualPred}, for a design of the state-based scheduler, and the corresponding optimal controller and observer. There are two estimators in this architecture, and hence, we call it a dual predictor architecture \cite{Ramesh2009}. This architecture has been referred to previously in \cite{Xu2004a}, in the context of mobile networks. The scheduler, observer and controller blocks are described below.

\noindent \textbf{Scheduler ($\mathcal{S}$):} The scheduler output $\gamma_k$ is given by
\begin{equation}
\gamma_k = f(x_k,\hat{x}_{^{k|\tau_{k-1}}}) = \begin{cases}
1, & |x_k - \hat{x}_{^{k|\tau_{k-1}}}|^2 > \epsilon, \\
0, & \textrm{otherwise}.
\end{cases}
\label{Eq:InnoSched}
\end{equation}
Here, $\hat{x}_{^{k|\tau_{k-1}}}$ is the estimate at the controller at time $k$ if the current packet is not scheduled for transmission. To realize such a scheduling policy, the observer must be replicated within the scheduler, and for the observer to be able to subtract the applied control, the controller must also be replicated within the scheduler. An explicit ACK is required to realize this information pattern.

\noindent \textbf{Observer ($\mathcal{O}$):} The input to the observer is the signal $y_k = \delta_k x_k$. The observer generates the estimate $\hat{x}_{^{k|k}}$ as given by
\begin{equation} \label{Eq:Estimatekk_DP}
\hat{x}_{^{k|k}} = \bar{\delta}_k \hat{x}_{^{k|\tau_k}} + \delta_k x_k \; .
\end{equation}
Recall that $\bar{\delta}_k = 1-\delta_k$ takes a value $1$ when the packet is not transmitted. In such a case, the estimate is given by $\hat{x}_{^{k|\tau_k}}$, a model based prediction from the last received data packet at time $\tau_k$. This estimate is given by
\begin{equation} \label{Eq:Estimatektau_DP}
\hat{x}_{^{k|\tau_k}} = A \hat{x}_{^{k-1|k-1}} + Bu_{k-1} \; .
\end{equation}

\noindent \textbf{Controller ($\mathcal{C}$):} The controller generates the signal $u_k$ based on the estimate alone, as given by
\begin{equation}
\label{Eq:Controller_DP}
u_k = - L_k \hat{x}_{^{k|k}} \; ,
\end{equation}
where $L_k$ is defined in (\ref{Eq:uCE}).

Note that the scheduling criterion described in (\ref{Eq:InnoSched}) can be rewritten as
\begin{align*}
|x_k - \hat{x}_{^{k|\tau_{k-1}}}|^2 ={}& |Ax_{k-1} + Bu_{k-1} + w_{k-1} - A\hat{x}_{^{k-1|k-1}} - Bu_{k-1}|^2 \\
={}& |A \tilde{x}_{^{k-1|k-1}} + w_{k-1}|^2 = |\tilde{x}_{^{k|\tau_{k-1}}}|^2 \; .
\end{align*}
Here, we use $\hat{x}_{^{k|\tau_{k-1}}}$ as $\tau_k$ is not defined without $\delta_k$. The scheduling criterion $|\tilde{x}_{^{k|\tau_{k-1}}}|^2 \le \epsilon$ captures the per-sample variance of the estimation error, when no transmission is scheduled. Taking expectations on both sides, we get $\tr \{ P_{^{k|\tau_{k-1}}} \} \le \epsilon$. The scheduler attempts to threshold the variance of the estimation error, but this cannot be guaranteed in a network with multiple traffic sources. Also, note that the scheduling policy is a symmetric function of its arguments, as in Proposition~\ref{P:OptObserver}. We now state the main result of this section. \newline

\begin{theorem}
\label{T:DualPredState}
For the closed-loop system given by the plant in (\ref{Eq:StateSpace4}), the state-based dual predictor architecture in (\ref{Eq:InnoSched})--(\ref{Eq:Controller_DP}), and the cost criterion in (\ref{Eq:LQGCriterion4}), it holds that
\newcounter{Tcount}
\begin{list}{\roman{Tcount}.}
{\usecounter{Tcount}}
\item The estimate (\ref{Eq:Estimatekk_DP}) minimizes the mean-squared estimation error.
\item The control signal does not have a dual effect.
\item The Certainty Equivalence Principle holds and the optimal control law is given by (\ref{Eq:Controller_DP}).
\item The LQG cost is given by
\begin{align}
J_{\textrm{DP}} ={}& \hat{x}_0^T S_0 \hat{x}_0 + \tr\{ S_0 P_0 \} + \sum_{n=0}^{N-1} \tr\{ S_{n+1} R_w + (L_n^T (Q_2 + B^T S_{n+1} B) L_n) P_{^{n|n}} \} \; , \label{Eq:J_DP}
\end{align}
where $P_{^{k|k}}$ is the error covariance of the estimate at the observer, with $S_N = Q_0$ and $S_k$ obtained by backward iteration of (\ref{Eq:Iter_s_k}).
\end{list}
\end{theorem}

\begin{IEEEproof}
Evaluating the expression $\E[x_k|\Ick]$, we get
\begin{equation*}
\E[x_k|\Ick] = \begin{cases}
x_k, & \delta_k = 1, \\
\E[A^{k-\tau_k}x_{\tau_k} + \sum_{^{\ell=1}}^{_{k-\tau_k}} (A^{\ell-1} B u_{k-\ell} + A^{\ell-1} w_{k-\ell}) &  \\
\quad \quad |\delta_{\tau_k} = 1, \vect{\delta}_{^{\tau_k+1}}^{_k} = 0, y_{\tau_k}=x_{\tau_k}, \vect{u}_{^0}^{_{k-1}}], & \delta_k = 0.
\end{cases}
\end{equation*}
Due to the use of a symmetric scheduling policy (\ref{Eq:Symf}), we know that
\begin{align*}
A^{k-\tau_k}x_{\tau_k} + \sum_{^{n=1}}^{_{k-\tau_k}} A^{n-1} B u_{k-n} &= A \hat{x}_{^{k-1|k-1}} + B u_{k-1} \; , \; \E[\sum_{^{\ell=1}}^{_{k-\tau_k}} A^{\ell-1} w_{k-\ell} | \vect{\delta}_{^{\tau_k+1}}^{_k} = 0] = 0 \; .
\end{align*}
We use the above equations to obtain
\begin{equation*}
\E[x_k|\Ick] = \begin{cases}
x_k, & \delta_k = 1, \\
A \hat{x}_{^{k-1|k-1}} + B u_{k-1}, & \delta_k = 0.
\end{cases}
\end{equation*}
Thus, the estimate in (\ref{Eq:Estimatekk_DP}) is the MMSE estimate \cite{Kailath2000}.

The error covariance at the estimator is given by (\ref{Eq:DualP}), where, from (\ref{Eq:InnoSched}) and (\ref{Eq:delta4}), it is clear that the scheduler outcome $\gamma_k$ and the MAC output $\delta_k$ do not depend on the applied controls $\vect{u}_{^0}^{_{k-1}}$. Thus, the error covariance satisfies the definition in (\ref{Eq:Def_DualEffect4}), and the control signal in this architecture does not have a dual effect.

From the above conclusion, note that the scheduling policy in (\ref{Eq:InnoSched}) is of the form (\ref{Eq:CheckSched}). Thus, from Corollary~\ref{C:CEnetwork}, we know that the optimal controller for this setup is certainty equivalent. Then, the optimal control signal is given by (\ref{Eq:uCE}), which has the same form as the controller in this architecture (\ref{Eq:Controller_DP}). The expression for the control cost remains the same as in the case with partial state information, and is given by (\ref{Eq:J_DP}).
\end{IEEEproof}

Thus, the dual predictor architecture results in a simplified design for the closed-loop system, which is optimal within its class (as seen from Corollaries~\ref{C:CE} and \ref{C:CEnetwork}), but not optimal among all possible architectures. 

%%%%%%%%%%%%%%%%%%%%%%%%%%%%%%%%%%%%%%%%%%%%%%%%%%%%%%%%%%%%%%%%%%%%%%%%%%%%%%%%

\newpage
\section{Extensions and Discussions} \label{S:ExtnDisc}

\IEEEPARstart{I}{n} this section, we extend the above results to an output-based system. We also identify the existence of a dual effect when the cost function penalizes network usage and when the transmission, with a state-based scheduler, occurs over limited data-rate channels. Finally, we discuss the dual effect property that we have encountered in this problem with respect to other NCS architectures.

\subsection{Measurement-based Scheduler} \label{SS:MeasSched}

We now consider a system without full state information, but with co-located measurements. We show that by placing an optimal observer, a Kalman Filter (KF) at the sensor, to estimate the state of the linear plant, and basing the scheduler decisions on this estimate, instead of on the measurement, we are able to re-establish the same problem formulation as before.

Consider a linear plant with a state $z_k$, and a measurement $m_k$ given by
\begin{equation} \label{Eq:z}
z_{k+1} = A z_k + B u_k + w_{z,k} \; , \quad m_k = C z_k + v_{z,k} \; ,
\end{equation}
where $w_{z,k}$ is i.i.d. zero-mean Gaussian with covariance matrix $R_{w,z}$. The initial state $z_0$ is zero-mean Gaussian with covariance matrix $R_{z,0}$. Also, the measurement $m \in \mathbb{R}^{m}$ and the matrix $C \in \mathbb{R}^{n x m}$. The measurement noise $v_{z,k}$ is a zero mean i.i.d Gaussian process with covariance matrix $R_{v,z} \in \mathbb{R}^{m x m}$, and it is independent of $w_{z,k}$.

We can always place a Kalman Filter at the sensor node, which receives every measurement $m_k$ from the sensor and updates its estimate ($\hat{z}^s_{^{k|k}}$) as
\begin{equation}
\label{eq:KFsensor}
\hat z^s_{^{k|k}} = A \hat z^s_{^{k-1|k-1}} + B u_{k-1} + K_{f,k} e_k \; ,
\end{equation}
where, $K_{f,k}$ denotes the gain of the Kalman filter and $e_k$ denotes the innovation in the measurement. The innovation can be shown to be Gaussian with zero-mean and covariance $R_{e,k}$. The error covariances for the predicted estimate and the filtered estimate are denoted $P^s_k$ and $P^s_{^{k|k}}$ respectively. These terms are given by
\begin{align*}
e_k &= m_k - C (A \hat z^s_{^{k-1|k-1}} + B u_{k-1}) \; ,  K_{f,k} = P^s_k C^T R_{e,k}^{-1} \; , R_{e,k} = C P^s_k C^T + R_{v,z} \; , \\
P^s_k &= A P^s_{^{k-1|k-1}} A^T + R_{w,z} \; , P^s_{^{k|k}} = P^s_k - K_{f,k} R_{e,k} K_{f,k}^T \; .
\end{align*}
Now, if we use the estimate to define a new state, such that $x_k \triangleq \hat{z}^s_{^{k|k}}$, we have a linear plant disturbed by i.i.d Gaussian process noise $w_k = K_{f,k} e_k$. Thus, we have re-established the problem setup from section \ref{S:ProbSetup4}, and the results from before can be applied to this plant. Note that the scheduler is now defined with respect to the estimate $\hat{z}^s_{^{k|k}}$ and not the measurements $m_k$.

\subsection{Penalizing Network Usage} \label{SS:PenalizingNW}
We have shown, in the proofs of Theorem~\ref{T:DualEffect_nNW} and Theorem~\ref{T:NEquiSys}, that the applied controls play a significant role in a state-based scheduler and cannot be removed from the scheduler inputs to create an equivalent setup without a dual effect. However, the minimizing solution to a cost criterion can render the effect of the applied controls redundant. To see an example of this, consider the problem of finding the jointly optimal scheduler-controller pair for the classical LQG cost criterion in (\ref{Eq:LQGCriterion4}). Since there is no penalty on using the network, the optimal scheduler policy is to transmit all the time. Now, the structure of the closed-loop system does not resemble the one presented in Theorem~\ref{T:DualEffect_nNW}, and consequently, that result does not hold. In this scenario, there is no incentive for the controller to influence the transmissions and the jointly optimal scheduler-controller pair $(f^{\mathds{1}},g^{\mathds{1}})$ is given by
\begin{equation}
\label{Eq:jointSC_lqg}
\begin{aligned}
f^{\mathds{1}}: \; \delta^{\mathds{1}}_k &= 1 \; \forall \; k \; , \quad g^{\mathds{1}}: \; u^{\mathds{1}}_k = -L_k x_k \; \forall \; k \; ,
\end{aligned}
\end{equation}
where $L_k$ is given in (\ref{Eq:uCE}). Note that in the rest of this paper, we do not consider finding the jointly optimal scheduler-controller pair, as the use of a contention-based MAC does not permit us to choose the schedule sequence.

Now, consider a cost criterion which penalizes the use of the network, such as
\begin{equation} \label{Eq:CostCriterion_lqgNw}
J_{\Lambda} = \min_{_{\vect{u}_{^0}^{_{N-1}},\vect{\delta}_{^0}^{_{N-1}}}} \E\left[ x_N^T Q_0 x_N + \sum_{s=0}^{N-1} \bigg(x_s^T Q_1 x_s + u_s^T Q_2 u_s + \Lambda \delta_s \bigg) \right ] \; ,
\end{equation}
where $Q_0$,$Q_1$ and $Q_2$ are positive definite weighting matrices and $\Lambda > 0$ is the cost of using the network. The optimal state-based scheduling policy chooses a schedule in relation to the penalty $\Lambda$, such that the average network use, i.e., $\E[\delta_k]$, decreases as $\Lambda$ increases. Thus, we state the following result.

\begin{lemma} \label{L:NwUsePenalizingCost}
For the closed-loop system defined by (\ref{Eq:StateSpace4})--(\ref{Eq:Controller4}), with no exogenous network traffic, the control signals derived from the jointly optimal scheduler-controller pair, which minimize the cost criterion in (\ref{Eq:CostCriterion_lqgNw}), exhibit a dual effect of order $r=2$.
\end{lemma}

\begin{IEEEproof}
It is easy to show that the scheduler-controller pair $(f^{\mathds{1}},g^{\mathds{1}})$ does not minimize the cost in (\ref{Eq:CostCriterion_lqgNw}). Now, the scheduler uses the policy in (\ref{Eq:StateSched4}) to select packets to send across the network. Thus, the closed-loop system has the same structure as in Theorem~\ref{T:DualEffect_nNW}, and there is a dual effect of order $r=2$ for any control signal in this setup.
\end{IEEEproof}

This provides the controller an incentive to modify the transmission outcome. As a result, the optimal scheduler and controller designs in this problem are coupled. Using the results of Lemma~\ref{L:DualEffect_withNetwork}, the above results can be extended to include the effect of exogenous network traffic.

\subsection{Using a rate constrained channel}
Our proof of the dual effect in Theorem~\ref{T:DualEffect_nNW} relies on the asymmetry in the resolution of the received information; the full state is sent with a transmission and only a single-bit quantized encoding of the state is sent when there is no transmission. However, data channels are generally rate-constrained, and a full state can never be sent. If the encoder-decoder pair on the sensor link uses $\mathbb{R}$ bits of information, the estimation error at the controller can be written as $\tilde{x}_{^{k|k}} = \delta_k \cdot (x_k - \E[x_k| \Ick, \delta_k=1]) + \bar{\delta}_k \cdot (x_k - \E[x_k| \Ick, \delta_k=0])$, in place of (\ref{Eq:DE_EstError}). Again, note that the estimation error is a function of $\delta_k$ due to the asymmetry in the number of bits in the received information with and without a transmission. Also, the applied controls cannot be removed from the above expression, unless the estimation error with and without a transmission result in the same expression, i.e., $x_k - \E[x_k| \Ick, \delta_k=1] = x_k - \E[x_k| \Ick, \delta_k=0]$. Hence, there is a dual effect with a state based scheduler, even while using a rate constrained channel for transmission.

\subsection{Relation to Other NCS Architectures}

The dual effect and certainty equivalence properties have been noted previously in other NCS problems. We discuss these occurrences and the connections to our problem setup below.

\begin{figure}
\begin{center}
\subfigure[Packet Losses ($d_k$)]{\label{Fig:nDE_PktLoss} \includegraphics[scale=0.52]{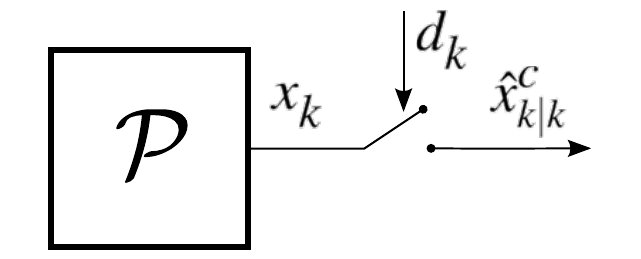}} \hspace{2mm}
\subfigure[Encoder-Decoder ($\mathcal{E}-\mathcal{D}$) Design]{\label{Fig:nDE_ED} \includegraphics[scale=0.52]{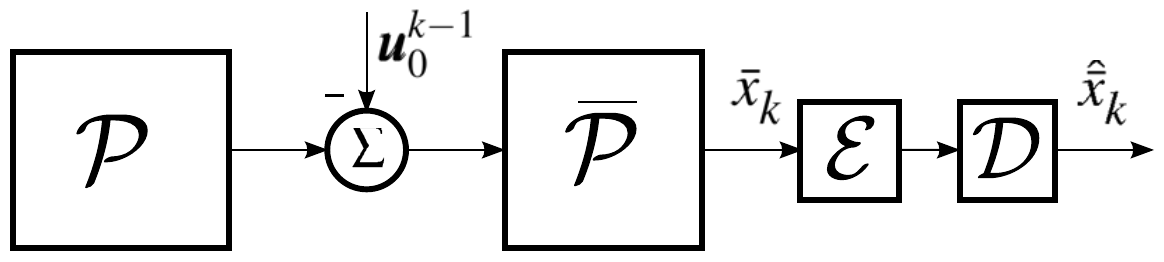}} \hspace{2mm}
\subfigure[State-based Scheduler]{\label{Fig:DualEffectInUnforcedx} \includegraphics[scale=0.52]{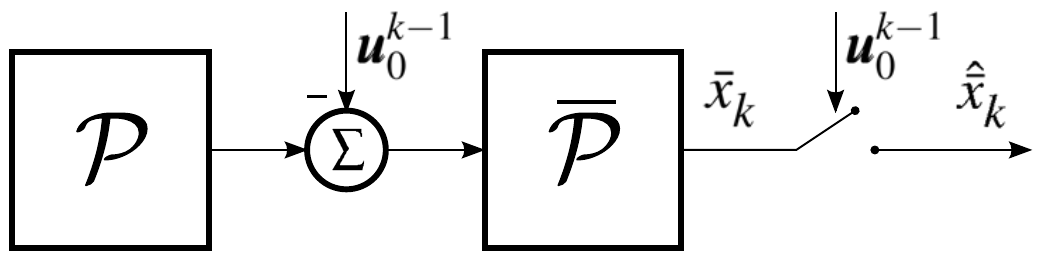}}
\caption[Dual effect and Certainty Equivalence in NCSs]{The estimate is not influenced by the applied controls in (a) and (b), with knowledge of the applied controls. In contrast, the applied controls cannot be removed from the decision process in (c). }
\end{center}
\vspace{-5mm}
\end{figure}

\begin{remark} [Packet Drops over a Lossy Network:]
Packet drops in a lossy network are not influenced by the applied controls (Fig.~\ref{Fig:nDE_PktLoss}). Hence, certainty equivalence holds, when there are packet drops on the sensor link \cite{Gupta2007}. However, when there are packet drops on the actuator link, separation holds only if there is an ACK of packets received or lost \cite{Schenato2007}.
\end{remark}

\begin{remark} [Importance of Side Information:]
In any NCS problem, the classical information pattern must be reconstructed for the Certainty Equivalence Principle to hold \cite{Witsenhausen1968}. This may require one or more explicit side information channels to convey acknowledgements of received packets back to the transmitters \cite{Bao2010,Schenato2007}.
\end{remark}

\begin{remark} [Encoder Design over Limited Data Rate Channels:]
In this problem, the encoder output is the only measurement available across the channel, which always contains the same number of information bits. Thus, the applied controls can be removed from the decision process, and they do not influence the estimation error, as shown in \cite{Tatikonda2004}.
\end{remark}

\begin{remark} [Event-based Systems:]
The results we have encountered in this paper show that the applied controls can push the state across the scheduler threshold, and influence the transmission outcome, as illustrated in Fig~\ref{Fig:DualRole}. This is a consequence of the unequal information in the measurement $y_k$, with and without a transmission.

A similar problem with a cost function such as (\ref{Eq:CostCriterion_lqgNw}), has been dealt with in \cite{Molin2009,Molin2010}. They use a transformation similar to the one presented for the encoder design problem in \cite{Nair2007}. There are, however, subtleties in defining an equivalence class for a state-based scheduler. Using an equivalent scheduler need not result in an equivalent system, as shown in Theorem~\ref{T:NEquiSys}.
\end{remark}

The dual effect is visible in any control signal applied to the plant, not just the optimal one, as the control signal will always influence the estimation error, irrespective of whether it has been designed to do so or not. Also, the dual effect exists despite knowledge of the applied controls at the scheduler, and knowledge of the scheduling decisions at the controller. In this context, the dual effect can be best explained as a coupling between the control and scheduling policies.

%%%%%%%%%%%%%%%%%%%%%%%%%%%%%%%%%%%%%%%%%%%%%%%%%%%%%%%%%%%%%%%%%%%%%%%%%%%%%%%%

\section{Examples} \label{S:Examples}
\IEEEPARstart{W}{e} present three examples in this section. The first example describes the problem setup, and illustrates the motivation for the problem. The second example illustrates the results of Theorem~\ref{T:DualEffect_nNW} and Theorem~\ref{T:NEquiSys}, which identify the dual role of the applied controls towards the information available to the controller. The third example illustrates the dual predictor architecture and provides an example of network-aware event triggering.
\begin{table}[tbp]
\vspace{-5mm}
\begin{center}
\caption[Control Cost Comparison for Example \ref{Example:Ssched1}]{A comparison of control costs with ($J_{\textrm{SS}}$) and without ($J_{\textrm{CN}}$) a state-based scheduler in the closed-loop, from Example \ref{Example:Ssched1}}
\label{Tb:ExampleSS1_Jlqg}
\begin{tabular}{| c || c | c | c |}
\hline Plant Type & $\mathcal{P}^{[T1]}$ & $\mathcal{P}^{[T2]}$ & $\mathcal{P}^{[T3]}$ \\ \hline \hline
$J_{\textrm{CN}}$ & $45.3074$ & $10.0028$ & $6.1213$ \\
$J_{\textrm{SS}}$ & $23.5785$ & $8.3489$ & $5.3803$ \\
\hline
\end{tabular}
\end{center}
\end{table}

\subsection{An Example of a Multiple Access NCS} \label{SS:ExampleSetup}

This example illustrates the role of a state-based scheduler in our problem formulation in Section~\ref{S:ProbSetup4}, where a number of closed-loop systems share a contention-based multiple access network on the sensor link. We use a $p$-persistent CSMA protocol in the MAC. The observer and controller are chosen for simplicity of design, not as optimizers of any cost. We look at the performance of this network of control loops, with and without the state-based scheduler.

\begin{example} \label{Example:Ssched1}
We consider a heterogenous network of $20$ scalar plants, indexed by $j \in \{1,\dots,20\}$. There are three different types of plants, $\mathcal{P}^{[T1]}$,$\mathcal{P}^{[T2]}$ and $\mathcal{P}^{[T3]}$, given by
\begin{equation} \label{Eq:StateSpace4_Ex1}
x^{(j)}_{k+1} = a^{[i]} x^{(j)}_k + u^{(j)}_k + w^{(j)}_k \; ,
\end{equation}
where $a^{[i]} \in \{1,0.75,0.5\}$, and $R^{[i]}_w \in \{1,1.5,2\}$, for the plant $\mathcal{P}^{[Ti]}$. The systems numbered $j \in \{1,\dots,6\}$ are of type $\mathcal{P}^{[T1]}$, $j \in \{7,\dots,13\}$ are of type $\mathcal{P}^{[T2]}$ and $j \in \{14,\dots,20\}$ are of type $\mathcal{P}^{[T3]}$. The plants are sampled with different periods given by $T^{[i]} \in \{10,20,25\}$, for the different types of plants, respectively. The state-based scheduler uses the criterion $x^{(j)^2}_k > \epsilon^{(j)}$. A $p$-persistent MAC, with synchronized slots, which permits three retransmissions is used. The persistence probability is given by $p^{(r)}_\alpha$, where $r$ denotes the retransmission index, and $p^{(r)}_\alpha = \{1,0.75,0.5\}$ for $r \in \{1,\dots,3\}$. The LQG criterion in (\ref{Eq:LQGCriterion4}), with $N = 10$ and $Q_0 = Q_1 = Q_2 = 1$ is used to design a certainty equivalent controller (\ref{Eq:uCE}) as an ad hoc policy, not an optimal one, as we know from Corollary~\ref{C:CEnetwork}. The observer calculates a simple estimate as given by (\ref{Eq:Estimatekk_DP})-(\ref{Eq:Estimatektau_DP}).

\begin{figure}[!t]
\begin{center}
\includegraphics*[scale=0.3,viewport=80 0 1000 625]{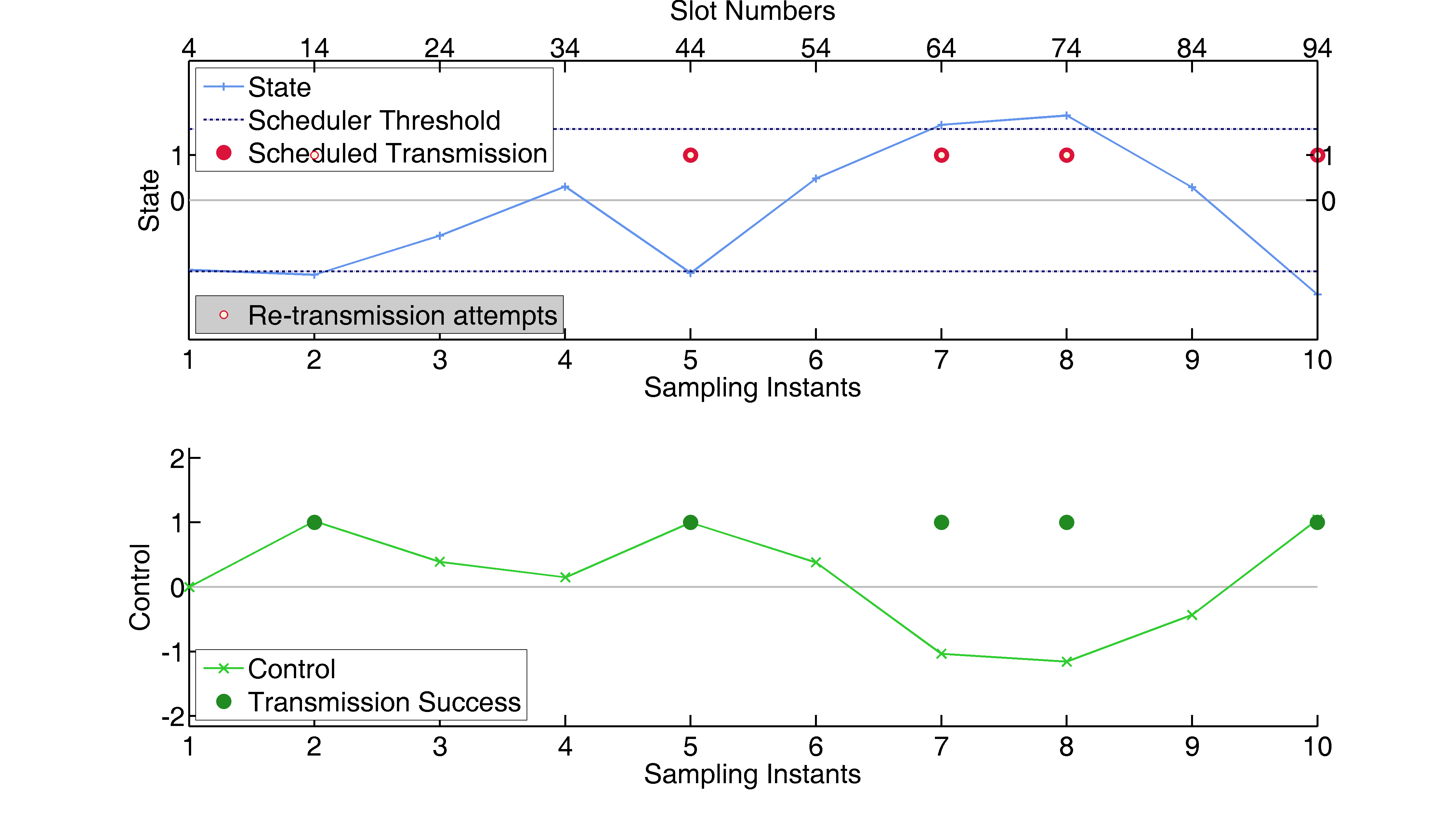} \vspace{-5mm}
\caption[Results of Example~\ref{Example:Ssched1}]{{\small The state and the control signal with the channel use pattern: the red circles denote transmission requests, the white circles denote MAC re-transmission attempts, and the green circles denote transmission success. Note that the requested bound on the state, which is marked with a dotted line, is sometimes exceeded due to network traffic. Also, the control signal corresponds closely to the state only when there is a successful transmission. }}
\label{Fig:XU1}
\vspace{-5mm}
\end{center}
\end{figure}

We look at the performance of a closed-loop system in this network without a state-based scheduler, i.e., when $\epsilon^{(j)} = 0$ for all $j$. The cost of controlling the plants in the contention-based network, is denoted $J^{[i]}_{\textrm{CN}}$, and the values are listed in Table~\ref{Tb:ExampleSS1_Jlqg}. We compare these to the costs obtained with a state-based scheduler in the closed-loop system, denoted $J^{[i]}_{\textrm{SS}}$, when $\epsilon^{(j)} = 2.5$. There is a marked improvement with a state-based scheduler in the closed-loop. Fig~\ref{Fig:XU1} depicts the state and the control signal for the first plant in this network. The above improvement is obtained due to fewer collisions in the contention-based MAC. The non-zero scheduling threshold reduces the traffic in the network, and increases the probability of a successful transmission for all the plants in the network.
\end{example}
\vspace{-7mm}

\subsection{A $2$-Step Horizon Example} \label{SS:Example2H}
We now look at a simple example to see the computational difficulties in identifying optimal estimates and controls for a system with a state-based scheduler in the closed-loop. We also show that for an equivalent scheduler such as $\tilde{f}$ in Section~\ref{SSS:EquiClass}, which renders the control signal free of a dual effect, the entire plant is altered, so the equivalence construction does not work.

\begin{example} \label{Example:Ssched2}
Consider a scalar plant, given by $x_{k+1} = ax_k + bu_k + w_k$, with $a,b \in \mathbb{R}$ and $x_0,w_k \sim \mathcal{N}(0,1)$. The scheduling law is given by
\begin{equation*}
\delta_k = \begin{cases}
1 & x_k \ge 0.5 \\
0 & \text{ otherwise}
\end{cases} \; .
\end{equation*}
Our aim is to find both the optimal controller, with dual effect, and the certainty equivalent controller for the equivalent scheduler and show that these result in different control actions for the same scheduling sequence. The controllers are designed to minimize the LQG cost (\ref{Eq:LQGCriterion4}), for a horizon of two steps, i.e., $N=2$, and with $Q_0,Q_1,Q_2 \in \mathbb{R}$. We first derive the optimal controller with dual effect. Then, for the same schedule, we define the certainty equivalent controller, assuming that an equivalent
scheduler of the form $\tilde{f}$ in (\ref{Eq:TildeSched}) has been designed. We compare the resulting control actions, and comment on the differences.

\noindent \textbf{Estimator: } The estimates $\hat{x}_{0|0}$ and $\hat{x}_{1|1}$ are obtained using (\ref{Eq:Estimate}). The estimation error covariances $P_{0|0}$ and $P_{1|1}$ are presented in the Appendix. Since the estimation error is non-Gaussian, we need to derive the probability density functions of the estimation errors at each time instant. This makes the computation of the estimation errors and the error covariances hard.

\noindent \textbf{Optimal Controller: } To solve for the optimal control signals, we use $V_1$ and $V_0$ from (\ref{Eq:Bellman4}). The complete derivations of $V_1$ and $V_0$ are presented in \cite{Ramesh2011}. We find the control signal $u_1$ that minimizes $V_1$, and get
\begin{equation}
u_1 = -\frac{abQ_0}{Q_2+b^2Q_0}\hat{x}_{1|1} \; . \label{Eq:u1opt}
\end{equation}
Then, to find $u_0$, we take a partial derivative of the expression for $V_0$ with respect to $u_0$ and get
\begin{equation}
\frac{\partial V_0}{\partial u_0} = 2u_0 (Q_2+b^2S_1) + 2\hat{x}_{0|0}abS_1 + \frac{a^2Q_0^2b^2}{Q_2+b^2Q_0} \cdot \frac{\partial}{\partial u_0} \left(
\E[P_{1|1}|\I^{^C}_{0}] \right) = 0 \; .
\label{Eq:u0opt}
\end{equation}
This can be simplified using the expression for $P_{1|1}$. When $\delta_0 = 1$, we have
\begin{equation*}
\frac{\partial V_0}{\partial u_0} = 2u_0 (Q_2+b^2S_1) + 2\hat{x}_{0|0}abS_1 - \frac{a^2Q_0^2b^2}{Q_2+b^2Q_0} b (w_{0,max}-\bar{w}_0)^2 \phi_{w_0}(w_{0,max}) = 0 \; ,
\end{equation*}
where $w_{0,max} = 0.5-ax_0-bu_0$. The final equation is obtained using Leibnitz rule. For the case when $\delta_0 = 0$, we have
\begin{equation*}
\frac{\partial V_0}{\partial u_0} = 2u_0 (Q_2+b^2S_1) + 2\hat{x}_{0|0}abS_1 - \frac{a^2Q_0^2b^2}{Q_2+b^2Q_0} b (e_{max}-\bar{e}_{\delta0})^2 \phi_{e_{\delta0}}(e_{max}) = 0 \; ,
\end{equation*}
where $e_{max} = 0.5-bu_0$ and again, Leibnitz rule was used. Solving these equations give the optimal $u_0$ for $\delta_0=1$ and $0$, respectively.

\noindent \textbf{CE Controller: } For the same scheduler outcomes $\delta_0$, $\delta_1$ obtained through an equivalent scheduler which has no dual effect, the certainty equivalent controller gives us the control signals
\begin{equation}
\begin{aligned} \label{Eq:u2CE}
u_1 &= -\frac{AbQ_0}{Q_2+b^2Q_0}\hat{x}_{1|1} \; , \\
u_0 &= -\frac{AbS_1}{Q_2+b^2S_1}\hat{x}_{0|0} \; .
\end{aligned}
\end{equation}
Note that the $u_1$ is found by minimizing $V_1$, which results in the same expression as for the optimal controller (\ref{Eq:u1opt}). However, $u_0$ for the CE controller is obtained by solving the equation
\begin{equation} \label{Eq:Solveu0CE}
2u_0 (Q_2+b^2S_1) + 2\hat{x}_{0|0}abS_1 = 0 \; .
\end{equation}

\noindent \textbf{Discussion: } A comparison of the control signals for the CE controller (\ref{Eq:u2CE}) with $u_1$ and $u_0$ obtained in (\ref{Eq:u1opt}) and (\ref{Eq:u0opt}), shows that the signal $u_1$ remains the same. However, $u_0$ is different, and displays a dual effect in the optimal controller. From (\ref{Eq:Solveu0CE}), it is clear that the additional term in (\ref{Eq:u0opt}) alters the solution for the optimal controller.

This observation can be explained as follows. In a controller with a dual effect, the control signal can be chosen to probe the plant state in order to improve the quality of the estimate. However, there is no motive in improving the estimate in a one-step optimization process. Thus, $u_1$ is the same for both controllers. When the optimization is performed over two steps, a probing effect in the first step can improve the estimate and the corresponding control applied in the next step. Thus, $u_0$ is different in the optimal controller for a state-based scheduler.

This example shows us that even the same schedule can result in a different control sequence for a system without a dual effect. Thus, an equivalent construction for the scheduler does not result in an equivalent system in our setup.
\end{example}

\begin{figure}[tb]
\hspace{-1.5cm}
\begin{center}
\includegraphics[scale=0.25]{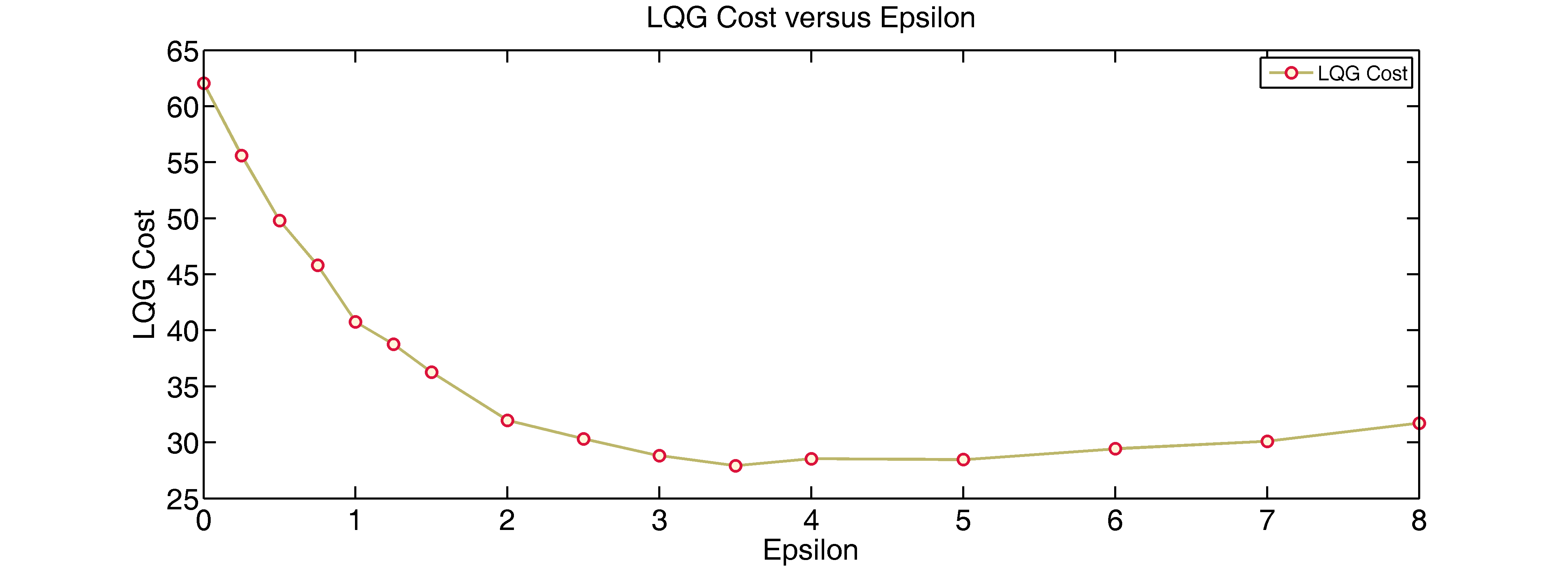}
\caption[Results of Example~\ref{Example:Ssched3}]{The control cost $J_{\textrm{DP}}$ versus the scheduler threshold $\epsilon$. For low thresholds, the high traffic in the network causes collisions, and a high $J_{\textrm{DP}}$. Very high values of $\epsilon$ result in an under-utilized network, and a high $J_{\textrm{DP}}$ due to insufficient transmissions.} \label{Fig:ExampleSS3_Jdp}
\end{center}
\vspace{-5mm}
\end{figure}

\subsection{An Example of the Dual Predictor Architecture} \label{SS:ExampleDP}
In this example, we present the dual predictor architecture, as applied to a shared network. We tune the threshold of the state-based scheduling law to probabilistically guarantee an achievable control performance, given the traffic over the network. We use a homogenous network in this example to simplify the comparison of control cost versus the scheduling threshold.

\begin{example} \label{Example:Ssched3}
We consider a shared network of $20$ scalar plants, indexed by $j \in \{1,\dots,20\}$ and given by (\ref{Eq:StateSpace4_Ex1}), where $a^{(j)} = 1$ and $R^{(j)}_w = 1$ for all $j$. The plants are sampled with a period given by $T = 10$. The innovations-based scheduler uses a similar criterion to (\ref{Eq:InnoSched}), where $\epsilon$ is the threshold of the scheduler. A $p$-persistent MAC, with synchronized slots, which permits three retransmissions is used. The persistence probability is given by $p^{(r)}_\alpha$, where $r$ denotes the retransmission index and $p^{(r)}_\alpha = \{1,0.75,0.5\}$ for $r \in \{1,\dots,3\}$. The LQG criterion in (\ref{Eq:LQGCriterion4}), with $N = 10$ and $Q_0 = Q_1 = Q_2 = 1$ is used to design the optimal certainty equivalent controller (\ref{Eq:uCE}). The observer calculates the MMSE estimate given by (\ref{Eq:Estimatekk_DP})-(\ref{Eq:Estimatektau_DP}).

The effect of varying $\epsilon$ on the control cost is shown in Fig.~\ref{Fig:ExampleSS3_Jdp}. For very high values of $\epsilon$, the network is under-utilized, and almost all the transmissions are successful. However, the control cost is high as the number of transmissions is low. As we decrease $\epsilon$, the control cost initially decreases due to increased use of the network. However, for very low values of $\epsilon$, the network is over-utilized and this results in collisions. Thus, the control cost increases again, due to dropped packets.

\begin{figure}[!b]
\vspace{-1cm}
\begin{center}
\includegraphics*[scale=0.3,viewport=100 0 1000 700]{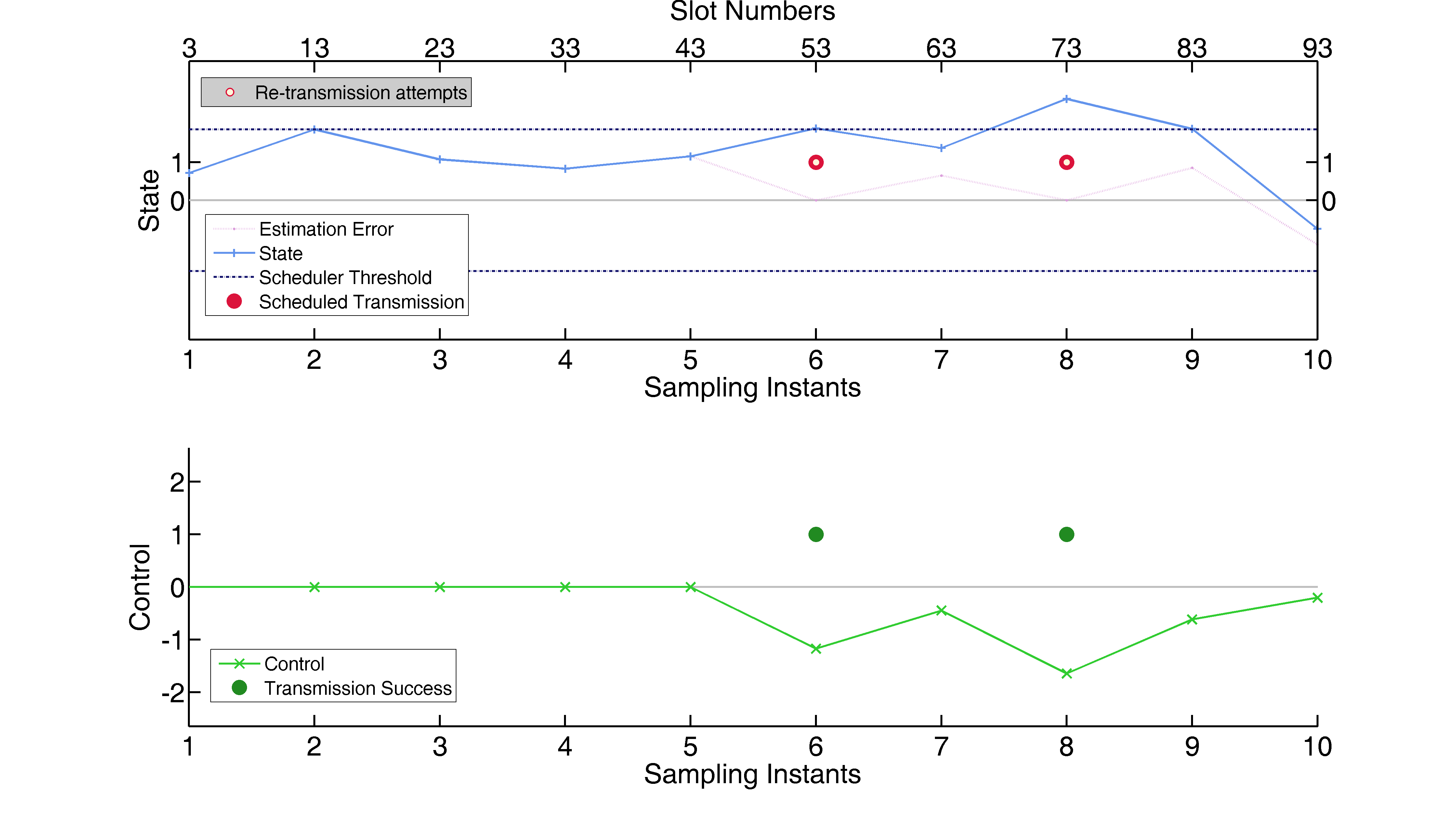} \vspace{-5mm}
\caption[Results of Example~\ref{Example:Ssched3}]{The estimation error, state and control signal with the channel use pattern. Note that the requested bound on the predicted estimation error, which is marked with a dotted line, is rarely exceeded. Also, the control signal corresponds closely to the state only when there is a successful transmission. }
\label{Fig:XU_DP}
\end{center}
\vspace{-5mm}
\end{figure}

Fig.~\ref{Fig:XU_DP} depicts the state and control signal of the first plant obtained from our simulation, for the best value of $\epsilon$ picked from the above plot. Note that the estimation error is bounded, with a probability of $0.94$, by the scheduling threshold, for the value $\epsilon^{(1)} = 3.5$, and the resulting control cost is $J_{\textrm{DP}} = 27.9235$.

It is interesting to note, in Fig.~\ref{Fig:ExampleSS3_Jdp}, that the cost function is quite flat. Thus, it is not very important to use the optimal scheduling threshold $\epsilon$.
\end{example}

%%%%%%%%%%%%%%%%%%%%%%%%%%%%%%%%%%%%%%%%%%%%%%%%%%%%%%%%%%%%%%%%%%%%%%%%%%%%%%%%
\section{ACKNOWLEDGMENTS}

The authors are grateful for the discussions with Maben Rabi, Lei Bao and Ather Gattami. Their insights and comments have helped to shape this work. This work was supported by the Swedish Research Council, VINNOVA (The Swedish Governmental Agency for Innovation Systems), the Swedish Foundation for Strategic Research, the Knut and Alice Wallenberg Foundation and the EU projects FeedNetBack and Hycon$2$.

%%%%%%%%%%%%%%%%%%%%%%%%%%%%%%%%%%%%%%%%%%%%%%%%%%%%%%%%%%%%%%%%%%%%%%%%%%%%%%%%

\section{Conclusions and Future Work} \label{S:ConclFW}

\IEEEPARstart{T}{his} paper investigates the effects of a state-based scheduler on the design of a closed loop system. We find that the optimal controller for a NCS with a state-based (or measurement-based) scheduler is, in general, difficult to find. This is due to the dual effect of the control signals in the given setup, wherein the controller has an incentive to push the state past the scheduler threshold and modify the estimation error across the network. This implies that the optimal scheduler, observer and controller designs are coupled. However, we identify a dual predictor architecture, which results in the desired separation in design of the scheduler, observer and controller. The scheduling function in this architecture is constrained to be a symmetric function of its arguments, such that the resulting schedule is not a function of the past applied controls.

Analyzing the performance of a network of systems using the dual predictor architecture is a challenging direction of work, for the future. Identifying interesting, if not optimal, control policies in the more general case of state-based schedulers with a dual effect, has been left for the future. A complete extension to include distributed measurements of the state is another task for the future.

%%%%%%%%%%%%%%%%%%%%%%%%%%%%%%%%%%%%%%%%%%%%%%%%%%%%%%%%%%%%%%%%%%%%%%%%%%%%%%%%

\appendix \label{App:Estimates_Ex2H}
\section*{Derivation of the $2$-Step Horizon Example}
Here are the expressions for the estimation error covariances in Example~\ref{SS:Example2H}. For a detailed derivation, refer \cite{Ramesh2011}.

The estimation error covariance at time $k=0$ is given by
\begin{equation} \label{Eq:ExSS2_P0}
P_{0|0} = \begin{cases}
0, & \delta_0 = 1, \\
R_{\tilde{x}_0}, & \delta_0 = 0,
\end{cases} \quad \text{where, } \;
\begin{aligned}
R_{\tilde{x}_0} &= \E[(x_0 - \bar{x}_{\delta0})^2|x_0<0.5] \\
&= \int_{-\infty}^{0.5-\bar{x}_{\delta0}} x^2 \phi_{x_{\delta0}}(x+\bar{x}_{\delta0}) dx \; ,
\end{aligned}
\end{equation}
where $\bar{x}_{\delta0} := \E[x_0|x_0<0.5] = \int_{-\infty}^{0.5} x \phi_{x_{\delta0}} (x) dx$, $\phi_{x_{\delta0}}$ is the conditional probability distribution function (pdf) of $x_0$, conditioned on $x_0<0.5$. Thus, $\phi_{x_{\delta0}}(x) = \phi_{x_0}(x)/Pr(x_0<0.5)$, where $\phi_{x_0}$ is the pdf of $x_0$. The probability of a non-transmission is given by $Pr(x_0<0.5) = \int_{-\infty}^{0.5} \phi_{x_0}(x) dx$.

Let us denote $e_1$ as the unknown part of $x_1$ before $y_1$ is received:
\begin{equation*}
e_1 =
\begin{cases}
w_0, & \delta_0=1, \\
ax_0 + w_0, & \delta_0 = 0,
\end{cases} \quad \text{and } \;
\quad \phi_e(\epsilon) =
\begin{cases}
\phi_{w_0}(\epsilon), & \delta_0=1, \\
\phi_{e_{\delta0}}(\epsilon), & \delta_0 = 0,
\end{cases}
\end{equation*}
where, $\phi_e$ is the pdf of $e_1$, $\phi_{w_0}$ is the pdf of $w_0$ and $\phi_{e_{\delta0}}(\epsilon) = \int_{-\infty}^{0.5} \phi_{x_{\delta0}}(x) \phi_{w_0}(\epsilon-ax) dx$. We denote $\tilde{e}_1$ as the error in estimating $e_1$ after $y_1$ arrives, and $\bar{e}_{\delta0} = \E[ax_0+w_0|x_0<0.5,ax_0+w_0<0.5-bu_0]$. Now, the estimation error variance $P_{1|1}$ is given by
\begin{equation} \label{Eq:ExSS2_P1}
P_{1|1} = \begin{cases}
0, & \delta_1 = 1, \\
R_{e_1}, & \delta_1 = 0,
\end{cases}
\end{equation}
where $R_{e_1} = \E[\tilde{e}_1^2|\delta_1=0]$ is given by
\begin{equation*}
R_{e_1} =
\begin{cases}
\int_{-\infty}^{^{0.5-ax_0-bu_0-\bar{w}_0}} w^2 \frac{\phi_{w_0}(w+\bar{w}_0)} {Pr(w_0<0.5-ax_0-bu_0)} dw, &
\delta_0 = 1, \\
\int_{-\infty}^{^{0.5-bu_0-\bar{e}_{\delta0}}} \epsilon^2 \frac{\phi_{\delta0}(\epsilon+\bar{e}_{\delta0})}
{Pr(e_1<0.5-bu_0)} d\epsilon, & \delta_0 = 0.
\end{cases}
\end{equation*}
Note that increasing $u_0$ will decrease $R_{e_1}$.

%%%%%%%%%%%%%%%%%%%%%%%%%%%%%%%%%%%%%%%%%%%%%%%%%%%%%%%%%%%%%%%%%%%%%%%%%%%%%%%%
\bibliographystyle{ieeetr}
\bibliography{SSchedNCS}

\end{document}

%% file: crenv.tex
\usepackage{amsmath} % assumes amsmath package installed
\usepackage{amssymb}  % assumes amsmath package installed
\usepackage{bm}  % bold math, preferred over amsbsy
\usepackage{dsfont}
\usepackage{epsfig} % for postscript graphics files
\usepackage{color}
\usepackage{subfigure}

\def\I{{\mathds{I}}}
\def\Ick{{\mathds{I}^{^C}_k}}

\DeclareMathOperator*{\E}{{\mathbb E}}        % expectation
\DeclareMathOperator{\tr}{tr}

\newcommand{\vect}[1]{\bm{#1}} % bold for vectors
\newtheorem{theorem}{Theorem}[section]
\newtheorem{corollary}[theorem]{Corollary}
\newtheorem{lemma}[theorem]{Lemma}
\newtheorem{proposition}[theorem]{Proposition}

\newtheorem{definition}{Definition}[section]

\newenvironment{example}[1][Example]{\begin{trivlist}
\item[\hskip \labelsep {\bfseries #1}]}{\end{trivlist}}
\newenvironment{remark}[1][Remark]{\begin{trivlist}
\item[\hskip \labelsep {\bfseries #1}]}{\end{trivlist}}

\newcommand{\qed}{\nobreak \ifvmode \relax \else \ifdim\lastskip<1.5em \hskip-\lastskip \hskip1.5em plus0em minus0.5em \fi \nobreak \hfill \vrule height0.5em width0.5em \fi}